\pdfoutput=1
\documentclass[sigconf]{acmart}
\usepackage{bm}
\usepackage{multirow}
\usepackage{subcaption}
\usepackage{bm}
\usepackage{adjustbox}
\usepackage{color}
\usepackage{diagbox}
\usepackage{url}
\usepackage[ruled, lined, linesnumbered, commentsnumbered, longend]{algorithm2e}

\AtBeginDocument{%
  \providecommand\BibTeX{{%
    \normalfont B\kern-0.5em{\scshape i\kern-0.25em b}\kern-0.8em\TeX}}}

\copyrightyear{2021}
\acmYear{2021}
\setcopyright{acmlicensed}
\acmConference[MM '21]{Proceedings of the 29th ACM International Conference on Multimedia}{October 20--24, 2021}{Virtual Event, China}
\acmBooktitle{Proceedings of the 29th ACM International Conference on Multimedia (MM '21), October 20--24, 2021, Virtual Event, China}
\acmPrice{15.00}
\acmDOI{10.1145/3474085.3475416}
\acmISBN{978-1-4503-8651-7/21/10}

\begin{document}
\fancyhead{}
\title{Self-supervised Consensus Representation Learning for Attributed Graph}



\author{Changshu Liu}
\affiliation{
 \institution{School of Computer Science and Engineering, 
 University of Electronic Science and Technology of China
    }
 }
 \email{ericliu9866@gmail.com}
\author{Liangjian Wen}
\affiliation{
\institution{
The Noah's Ark Lab,
Huawei Technologies Company Limited
}
}
\email{wenliangjian1@huawei.com}

\author{Zhao Kang}\authornote{Corresponding author}
\affiliation{
 \institution{School of Computer Science and Engineering, 
 University of Electronic Science and Technology of China
    }
 }
 \email{zkang@uestc.edu.cn}
 
 \author{Guangchun Luo}
 \affiliation{
 \institution{School of Information and Software
Engineering,
 University of Electronic Science and Technology of China
    }
 }
 \email{gcluo@uestc.edu.cn}
 
 \author{Ling Tian}
 \affiliation{
 \institution{School of Computer Science and Engineering, 
 University of Electronic Science and Technology of China
    }
 }
 \email{lingtian@uestc.edu.cn}






\renewcommand{\shortauthors}{Trovato and Tobin, et al.}

\begin{abstract}
Attempting to fully exploit the rich information of topological structure and node 
features for attributed graph, we introduce self-supervised learning
mechanism to graph representation learning and propose a novel Self-supervised Consensus 
Representation Learning (SCRL) framework.
In contrast to most existing works that only explore one graph, our proposed 
SCRL method treats graph from two perspectives: topology graph and feature graph.
We argue that their embeddings should share some common information, 
which could serve as a supervisory signal. 
Specifically, we construct the feature graph of node features via 
k-nearest neighbour algorithm. Then graph convolutional network (GCN) encoders extract features from two graphs
respectively. Self-supervised loss is designed to maximize the agreement of 
the embeddings of the same node in the topology graph and the feature graph. 
Extensive experiments on real citation networks and social networks demonstrate 
the superiority of our proposed SCRL over the state-of-the-art methods on
semi-supervised node classification task. Meanwhile, compared with its main competitors,
SCRL is rather efficient. The source code is available at \url{https://github.com/topgunlcs98/SCRL}.
\end{abstract}




\maketitle

\section{Introduction}
With the proliferation of information collected on the Internet, besides topological structure, vertices in a graph are often associated with 
content information, i.e., node attributes, on the top of a plain graph.
This type of system is modeled by attributed graph \cite{lin2021graph}.
It poses a new challenge:
how to learn a holistic representation for attributed graph?
Since deep learning methods can 
effectively extract useful representation of data, graph
neural networks have been applied on a wide range of disciplines, including 
social network \cite{qiu2018deepinf}, chemistry and biology \cite{duvenaud2015convolutional, rhee2017hybrid},
traffic prediction \cite{cui2019traffic, rahimi2018semi}, text classification
 \cite{hamilton2017inductive, defferrard2016convolutional}, and
 knowledge graph \cite{hamaguchi2017knowledge, wang2018cross}.
As a representative method in graph neural networks, GCN \cite{kipf2016semi} has shown impressive performance 
for semi-supervised classification task because it propagates feature information over the graph 
topology, providing a new fusion strategy for topological structure and node features.
By contrast, some models, like multi-layer perceptron (MLP) \cite{pal1992multilayer}, rely exclusively on node features.

\begin{figure*}[!htbp]
  \centering
  \includegraphics[width=0.9\textwidth]{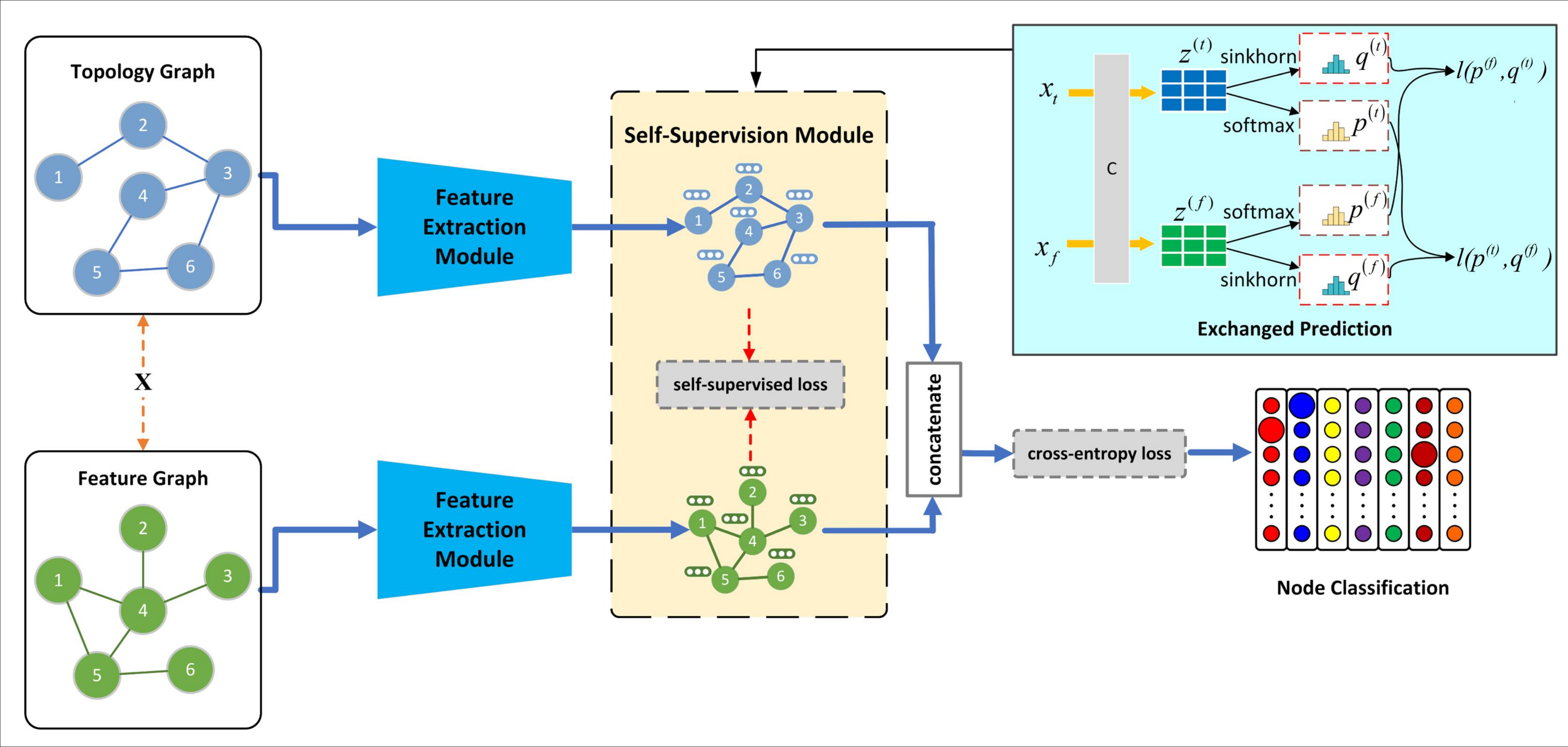}
  \caption{The framework of SCRL model. SCRL consists of two parts:
  feature extraction module and self-supervision module. First, we use node features to 
  construct feature graph. Then, in feature extraction module, we adopt two independent GCNs
  to extract the latent features of nodes in two graphs, respectively. Consensus representation is 
  learnt in the self-supervision module by solving the "exchanged prediction" problem. 
  We use cross-entropy loss to penalize the
  difference between prediction and the ground-truth label.}
  \label{fig:overview}
\end{figure*}
Some recent studies theoretically analyze the weakness of the fusion mechanism in GCN.
Li \emph{et al.} \cite{li2018deeper} indicate that the graph convolution of the GCN model is 
actually a special form of Laplacian smoothing on node features. 
However, repeatedly applying Laplacian smoothing may mix the features of vertices 
from different clusters and make them indistinguishable, which has a 
negative impact on downstream tasks \cite{ma2020towards}.
\cite{nt2019revisiting} and \cite{wu2019simplifying} show 
that GCN only performs low-pass filtering on feature vectors of nodes and
graph structure only provides a way to denoise the data. 
To further prove that the fusion capability of current GCN \cite{kipf2016semi} is not optimal, 
Wang \emph{et al.} \cite{wang2020gcn} set up a series of experiments and demonstrate that  
MLP \cite{pal1992multilayer} and DeepWalk \cite{perozzi2014deepwalk} can easily perform
better than GCN even under some simple situations that the 
correlation between node labels and features or topology structure is very obvious. 
The situation becomes more complex in reality.
For example, in a social network, the relationship between people is very complicated, 
while most of current GCN-based methods mainly consider the relationship between connected individuals and ignore other implicit information. In a citation network where nodes represent 
papers and there is an edge between two papers if they share the same author, if we conduct
information aggregation based on the original topology structure, it will ignore the situation
that an author may write papers belonging to different categories or similar papers are actually
written by different authors, hence we may mistakenly aggregate different types of papers together.

Towards a better fusion strategy integrating both node features and topology structure,
attention mechanism is also introduced. 
Graph attention network (GAT) \cite{velivckovic2017graph} assigns attention 
weights to every edge, which measures the local similarities among features of neighbors.  
Wang \emph{et al.} \cite{wang2020gcn} propose an adaptive 
multi-channel GCN (AMGCN) that learns suitable weights 
for fusing feature and topology information.
However, these attention-based approaches need to compute weights for every node, 
which is inefficient for large-scale graphs. These issues motivate us to design a new framework
which can improve the capability of fusing topology structure and node features and, at the same time,
enhance the computational efficiency.

In fact, there are correlations between graph structure and node 
features and the underlying information from these two aspects could supervise each other \cite{kang2020relation}.
For example, a blogger may tend to follow another one who has similar interests listed on his/her homepage.
Hence, we mine the common information between node features and graph structure
with an efficient self-supervised mechanism.

Specifically, we argue that the shared information could boost the performance of 
downstream tasks.
As shown in Fig.\ref{fig:overview}, firstly a kNN graph is generated from the
node features as the feature graph. Secondly, with the feature graph and topology
graph, we use different convolution modula to extract node embeddings in
feature space and topology space. Finally, in order to learn a consensus 
representation, we propose a self-supervised module that uses the node
embedding in one graph to predict the classification result of the same node in the other graph. 

Our main contributions are summarized as follows:
\begin{itemize}
    \item We propose a novel self-supervised framework to learn a consensus representation 
    for attributed graph. To the best of our knowledge,
    we are the first to explore the role of self-supervised 
    mechanism in fusing the topology information and node feature information of graph.
    \item We develop an efficient graph representation learning algorithm which takes less time in training 
    compared with other approaches.
    \item We conduct extensive experiments on a series of datasets. It shows
    that our approach outperforms many state-of-the-art methods on semi-supervised node
    classification task. Even with very small amounts of labeled data, our SCRL is still superior to main competitors.
\end{itemize}

\section{RELATED WORKS}
\subsection*{Graph-based Semi-supervised Learning}
Large numbers of approaches for semi-supervised learning on graphs have been
proposed in the past two decades. Some early studies use graph Laplacian
regularization and graph embedding \cite{kang2020structured,kang2020robust} to learn graph representation. Inspired by the Skip-gram model \cite{mikolov2013distributed}
for natural language processing, Perozzi \emph{et al.} propose DeepWalk \cite{perozzi2014deepwalk} 
that learns latent representation from sampled truncated random walks via the prediction of 
local neighborhood of nodes. Subsequently, some varients are proposed to improve 
DeepWalk, prominent examples include Line \cite{tang2015line}
 and node2vec \cite{grover2016node2vec}.
 
Recently, thanks to the breakthroughs in deep learning, attention has 
turned to various types of graph neural networks.
Bruna \emph{et al.} \cite{bruna2013spectral} propose a general graph convolution framework based on graph Laplacian.
\cite{defferrard2016convolutional} then optimizes it utilizing 
Chebyshev polynomial approximation to improve efficiency.
Besides, GCN \cite{kipf2016semi} uses a localized first-order approximation to
simplify the convolution operation. GAT \cite{velivckovic2017graph} gives 
different attention weights to different nodes in a neighborhood to aggregate node features.
Demo-Net \cite{wu2019net} builds a degree-specific graph neural network for 
both node and graph classification. MixHop \cite{abu2019mixhop} utilizes 
multiple powers of adjacency matrix to learn general mixing of neighborhood information.
However, these methods only use 
a single topology graph for node aggregation, which may cause graph structure
to be emphatically considered and fails to make full use of rich feature information.
AMGCN \cite{wang2020gcn} utilizes attention mechanism to merge embeddings extracted 
from topology graph and feature graph. However, attention-based methods always require 
high time and space complexity. 
\subsection*{Self-supervised Learning}
Self-supervised learning aims to learn representative features without label
information, which is able to reduce human cost for annotating data. 
Self-supervised learning has found many successful applications ranging from language modeling
 \cite{lample2019cross} to computer vision \cite{pathak2016context,lv2021pseudo}.
As a category of self-supervised learning, contrastive learning trains the network by
comparing the representations learnt from augmented samples. For example,
MoCo \cite{he2020momentum} and SimCLR \cite{chen2020simple} construct negative pairs and positive
pairs via data augmentation techniques and then compare them.
However, it becomes computationally expensive for large datasets. Another class is cluster-based 
approaches. For instance, Caron \emph{et al.} \cite{caron2020unsupervised} present a simplified
training pipeline by mapping features to cluster prototypes. 

Recently, there are a few works focusing on self-supervised learning in the domain of graph.
M3S \cite{sun2020multi} utilizes self-supervised learning approach to improve the
generalization performance of GCN.
Deep Graph Infomax (DGI) \cite{velickovic2018deep} drives local network embeddings to capture global structural information by maximizing local mutual information. 
Deep graph contrastive representation learning (GRACE) \cite{Zhu2020vf} is a graph contrastive representation learning framework by maximizing agreement at the node level.
Most graph contrastive learning approaches conduct random corruption on nodes and edges,  
which may bring noise into original graph data and degrade the learnt representation.
Nevertheless, whether self-supervised mechanism can improve the capability of GCN in fusing topological structure
and node features still remains unexplored.

\section{The proposed methodology}
\subsection{Notation}
An attributed graph can be represented as $G=\left\{ A, X\right\}$, 
where $A \in \mathbb{R}^{N\times N}$ is the adjacency matrix of $N$ nodes 
and $X \in \mathbb{R}^{N \times d}$ is the node feature matrix 
wherein every node is described by a vector with $d$ dimensions.
Each node belongs to one out of M classes.
As for $A$,
$A_{ij}=1$ represents that there is an edge between node $i$ and node $j$ while
$A_{ij}=0$ indicates that node $i$ and node $j$ are not connected. In our study,
we derivate the corresponding feature graph $\hat{G}=\left\{ \hat{A}, X\right\}$,
which shares the same $X$ with $G$, but has a different adjacency matrix.
Therefore, topology graph and feature graph refer to 
$G$ and $\hat{G}$ respectively.

\subsection{The Framework of SCRL}
As shown in Fig.\ref{fig:overview}, we use topology graph and feature graph to capture 
the underlying information in topology space and feature space.
Our model mainly contains two components: the feature abstraction module that uses GCN 
to extract features from graph, the self-supervision module that measures
the consistency between the representations learned from the topology graph and feature graph.
\subsubsection*{\textbf{Feature Graph}}
\ 
\newline
Merely propagating feature information over topology graph may hinder the fusion capability
of GCN under some circumstances \cite{wang2020gcn}.
A natural idea would be to complement topology graph by fully making use of the information
inside node features.
Therefore we introduce feature graph into our work.

To represent the structure of nodes in the feature space, we build a 
kNN graph \textbf{$\hat{G}$} based on the feature matrix $X$. To be precise, 
a similarity matrix \textbf{$S$} is computed using the cosine similarity formula:
\begin{equation}
    S_{ij} = \frac{x_{i}\cdot x_{j}}{|x_{i}| \cdot |x_{j}|}
\end{equation}
where $S_{ij}$ is the similarity between node feature $x_i$ and node feature $x_j$.
Then for each node we choose the top k nearest neighbors and
establish edges. In this way, we construct feature graph.

\subsubsection*{\textbf{Feature Extraction Module}}
\ 
\newline
To extract meaningful features from graphs, we adopt GCN that is 
comprised of multiple graph convolutional layers. With the input graph G, the
$(l+1)$-th layer's output $H^{(l+1)}$ can be represented as:
\begin{equation}
    H^{(l+1)} = ReLU(D^{-\frac{1}{2}}AD^{-\frac{1}{2}}H^{(l)}W^{(l)})
\end{equation}
where $ReLU$ is the Relu activation function ($ReLU(\cdot) = max(0, \cdot) $), 
$D$ is the degree matrix of $A$, $W^{(l)}$ is a layer-specific trainable weight matrix, 
$H^{(l)}$ is the activation matrix in the $l$-th layer and $H^{(0)} = X$.
In our study we use two GCNs to exploit the information in topology and feature space. 
The output is donated by $x_t=\left\{ x_{t1}, x_{t2}, \cdots, x_{tN}\right\}$ 
and $x_f=\left\{ x_{f1}, x_{f2}, \cdots, x_{fN}\right\}$, respectively.

\subsubsection*{\textbf{Self-Supervision Module}}
\ 
\newline
An important component of our framework is self-supervision module that is used to fuse
the representations learnt by feature extraction modules. To convert the 
learnt representations into vectors of class scores, the representations of the i-th node
$x_{ti}$ and $x_{fi}$ are followed by a \emph{prototype head C}: $\mathbb{R}^U \rightarrow \mathbb{R}^B$. 
$U$ refers to the dimension of $x_{ti}$ and $x_{fi}$. $B$ refers to the number of prototypes.
In \emph{prototype head C}, we store a set of prototype vectors: 
$\left\{c_1, \cdots, c_B\right\}$.
Every prototype vector projects the node feature to a prototype/cluster.  
By computing the dot product between $x_{ti}$ and $c_b$, 
we can get the class score of node $i$ corresponding to prototype $b$ in topology graph.
Specifically,
\begin{equation}
    z^{(t)}_i = C(x_{ti}) = x_{ti}^\top C
\end{equation}
\begin{equation}
    z^{(f)}_i = C(x_{fi}) = x_{fi}^\top C
\end{equation}
\emph{C} is a linear layer. Prototypes are initialized randomly and are learnt
along with the training process.

Then we compute the probability of assigning prototype $j$ to node representation $x_i$ by taking 
the softmax of its projection:
\begin{equation}
    p_{ij}^{(t)} = \frac{exp(\frac{1}{\tau}z_{ij}^{(t)})}{\sum_{j^{\prime}}^{B} exp(\frac{1}{\tau}z_{ij^{\prime}}^{(t)})}
\end{equation}

\begin{equation}
    p_{ij}^{(f)} = \frac{exp(\frac{1}{\tau}z_{ij}^{(f)})}{\sum_{j^{\prime}}^{B} exp(\frac{1}{\tau}z_{ij^{\prime}}^{(f)})}
\end{equation}
where $\tau$ is a temperature parameter. 
We use $p_i^{(t)}=\left\{ p_{i1}^{(t)}, \cdots, p_{iB}^{(t)} \right\}$ and
$p_i^{(f)}=\left\{ p_{i1}^{(f)}, \cdots, p_{iB}^{(f)} \right\}$ to denote 
the probabilities of the $i$-th node belonging to different prototypes in topology graph 
and feature graph.

A key question in our self-supervision module is to find a `target' for the projection of 
prototype vectors.
Inspired by previous works \cite{asano2019self}, we utilize pseudo label in this module. 
We cast pseudo label assignment 
problem as an optimal transport problem and compute the soft labels using the 
iterative Sinkhorn algorithm \cite{cuturi2013sinkhorn}. The outputs are denoted
by $q_i^{(t)}$ and $q_i^{(f)}$ for $z_{i}^{(t)}$ and $z_{i}^{(f)}$ respectively.

With $p_i^{(t)}$, $p_i^{(f)}$ and $q_i^{(t)}$, $q_i^{(f)}$, we set up the "exchanged prediction" problem.
We assume that both topology graph and feature graph should produce the same label.
Therefore the pseudo labels obtained from one graph are predicted using the other graph.
To be precise, for every node, it is our goal to minimize the cross entropy of two pairs of probabilities: 
$q_i^{(f)}$, $p_i^{(t)} $ and $q_i^{(t)}$, $p_i^{(f)}$.
The loss function of exchanged prediction can be defined as:
\begin{equation}
    l_{ss}^{(i)} = l(p_{i}^{(t)}, q_i^{(f)}) + l(p_{i}^{(f)}, q_i^{(t)})
    \label{equ: ss_loss}
\end{equation}
\begin{equation}
    l(p_{i}^{(t)}, q_i^{(f)}) = -\sum_{b=1}^B q_{ib}^{(f)}\log p_{ib}^{(t)}
\end{equation}
\begin{equation}
    l(p_{i}^{(f)}, q_i^{(t)}) = -\sum_{b=1}^B q_{ib}^{(t)}\log p_{ib}^{(f)}
\end{equation}

Summing this loss over all nodes leads to the following loss function 
for the "exchanged prediction" problem. 

\begin{equation}
\begin{aligned}
    L_{ss} &= \frac{1}{N}\sum_{i=1}^N l(p_{i}^{(t)}, q_i^{(f)}) + l(p_{i}^{(f)}, q_i^{(t)})\\
    &=-\frac{1}{N}\sum_{i=1}^N[\frac{1}{\tau}x_{ti}^\top Cq_i^{(f)}+\frac{1}{\tau}x_{fi}^\top Cq_i^{(t)}\\
    &-\log\sum_{b=1}^{B}\exp(\frac{x_{ti}^\top c_b}{\tau})-\log\sum_{b=1}^{B}\exp(\frac{x_{fi}^\top c_b}{\tau})]
\end{aligned}
\end{equation}
This loss function is jointly minimized with respect to prototypes C and the parameters of the GCN encoders 
that produce representations $x_t$ and $x_f$.  

\subsubsection*{\textbf{Node Classification}}
\ 
\newline
Ideally, $X_t$ and $X_f$ should be close to each other.
To preserve the information from feature graph and topology graph, $X_t$ and $X_f$ are concatenated as the consensus representation $R$.
Then we use $R$ for semi-supervised classification
with a linear transformation and a softmax function.
$Y^{\prime}$ is the prediction result and $Y^{\prime}_{ij}$ is the probability 
of node $i$ belonging to class $j$. $W$ and $a$ are weights and bias of the linear layer, respectively.
\begin{equation}
    Y^\prime=\operatorname{softmax}(W \cdot R + a)
\end{equation}
Suppose there are $T$ nodes with labels in the training set. We adopt cross-entropy to measure 
the difference between prediction label $Y^\prime_{ij}$ and ground truth label $Y_{ij}$:
\begin{equation}
    {L}_{ce}=-\sum_{i=1}^{T} \sum_{j=1}^{M} Y_{i j} \ln Y^\prime_{i j}
    \label{equ: ce_loss}
\end{equation}

Finally, by combining $L_{ss}$ and  $L_{ce}$,  
overall loss function of our SCRL model can be represented as:
\begin{equation}
    L = L_{ce} + L_{ss}
    \label{equ: oa_loss}
\end{equation}
The parameters of the whole framework are updated via backpropagation.
The consensus representation of the attributed graph can be learnt at the same time.
The detailed description of our algorithm is provided in 
Algorithm \ref{algo: algo}.

\begin{algorithm}
    \caption{The proposed algorithm SCRL}
    \label{algo: algo}
    \small
    \SetKwFunction{isOddNumber}{isOddNumber}
    \SetKwInput{Input}{Input}
    \SetKwInput{Output}{Output}

    \KwIn{Node feature matrix $X$; adjacency matrix $A$; 
    node label matrix $Y$; maximum number of iterations $\eta$;
    temperature parameter $\tau$}
    
    Compute the feature graph $\hat{G}$ according to $X$ by running kNN algorithm.
    
    \For{$it=0$ \KwTo $\eta$}{
    \tcc{Consensus representation learning}
    
    $x_t$ = GCN($G$) \tcp{embeddings of two graphs}
    $x_f$ = GCN($\hat{G}$)
    
    $z^{(t)}$ = prototype($x_t$)
    
    $z^{(f)}$ = prototype($x_f$)
    
    
    $p^{(t)}$ = softmax($z^{(t)}$ / $\tau$)
    
    $p^{(f)}$ = softmax($z^{(f)}$ / $\tau$)
    
    
    $q^{(t)}$ = sinkhorn($z^{(t)}$)
    
    $q^{(f)}$ = sinkhorn($z^{(f)}$)
    
    
    
    Calculate the overall loss with Equation(\ref{equ: oa_loss})
    
    Update all parameters of framework according to
    the overall loss
    }
    
    Predict the labels of unlabeled nodes based on the trained framework.
    
    \KwOut{Classification results $Y^{\prime}$}
\end{algorithm}
\begin{table}[t]
\renewcommand{\arraystretch}{.9}
\caption{The statistics of datasets.}
\label{table:dataset}
\begin{tabular}{ccccc}
\hline
\textbf{Datasets}    & \textbf{Nodes} & \textbf{Edges} & \textbf{Dimensions} & \textbf{Classes} \\ \hline
\textbf{Citeseer}    & 3327           & 4732           & 3703                & 6                \\
\textbf{PubMed}      & 19717          & 44338          & 500                 & 3                \\
\textbf{ACM}         & 3025           & 13128          & 1870                & 3                \\
\textbf{BlogCatalog} & 5196           & 171743         & 8189                & 6                \\
\textbf{UAI2010}     & 3067           & 28311          & 4973                & 19               \\
\textbf{Flickr}      & 7575           & 239738         & 12047               & 9                \\ \hline
\end{tabular}
\end{table}
\begin{table*}[t]
\caption{Node classification results($\%$). L/C refers to the number of labeled nodes per class.}
\label{table:result}
\renewcommand{\arraystretch}{.9}
\begin{tabular}{|c|cc|cc|cc|cc|cc|cc|}
\hline
Dataset       & \multicolumn{6}{c|}{\textbf{ACM}}                                                                                                 & \multicolumn{6}{c|}{\textbf{BlogCatalog}}                                                                                                              \\ \hline
L/C           & \multicolumn{2}{c|}{20}                   & \multicolumn{2}{c|}{40}                   & \multicolumn{2}{c|}{60}                   & \multicolumn{2}{c|}{20}                          & \multicolumn{2}{c|}{40}                          & \multicolumn{2}{c|}{60}                          \\ \hline
Label Rate           & \multicolumn{2}{c|}{1.98\%}                   & \multicolumn{2}{c|}{3.97\%}                   & \multicolumn{2}{c|}{5.95\%}                   & \multicolumn{2}{c|}{2.31\%}                          & \multicolumn{2}{c|}{4.62\%}                          & \multicolumn{2}{c|}{6.93\%}                          \\ \hline
Metrics       & \multicolumn{1}{c|}{ACC} & F1             & \multicolumn{1}{c|}{ACC} & F1             & \multicolumn{1}{c|}{ACC} & F1             & \multicolumn{1}{c|}{ACC} & F1                    & \multicolumn{1}{c|}{ACC} & F1                    & \multicolumn{1}{c|}{ACC} & F1                    \\ \hline
DeepWalk \cite{perozzi2014deepwalk}      & 62.69                    & 62.11          & 63.00                    & 61.88          & 67.03                    & 66.99          & 38.67                    & 34.96                 & 50.80                    & 48.61                 & 55.02                    & 53.36                 \\
LINE \cite{tang2015line}         & 41.28                    & 40.12          & 45.83                    & 45.79          & 50.41                    & 49.92          & 58.75                    & 57.75                 & 61.12                    & 60.72                 & 64.53                    & 63.81                 \\
ChebNet \cite{defferrard2016convolutional}      & 75.24                    & 74.86          & 81.64                    & 81.26          & 85.43                    & 85.26          & 38.08                    & 33.39                 & 56.28                    & 53.86                 & 70.06                    & 68.37                 \\
GCN \cite{kipf2016semi}          & 87.80                    & 87.82          & 89.06                    & 89.00          & 90.54                    & 90.49          & 69.84                    & 68.73                 & 71.28                    & 70.71                 & 72.66                    & 71.80                 \\
kNN-GCN \cite{wang2020gcn}      & 78.52                    & 78.14          & 81.66                    & 81.53          & 82.00                    & 81.95          & 75.49                    & 72.53                 & 80.84                    & 80.16                 & 82.46                    & 81.90                 \\
GAT \cite{velivckovic2017graph}         & 87.36                    & 87.44          & 88.60                    & 88.55          & 90.40                    & 90.39          & 64.08                    & 63.38                 & 67.40                    & 66.39                 & 69.95                    & 69.08                 \\
Demo-Net \cite{wu2019net}     & 84.48                    & 84.16          & 85.70                    & 84.83          & 86.55                    & 84.05          & 54.19                    & 52.79                 & 63.47                    & 63.09                 & 76.81                    & 76.73                 \\
MixHop \cite{abu2019mixhop}       & 81.08                    & 81.40          & 82.34                    & 81.13          & 83.09                    & 82.24          & 65.46                    & 64.89                 & 71.66                    & 70.84                 & 77.44                    & 76.38                 \\
GRACE \cite{Zhu2020vf}       & 89.04                    & 89.00          & 89.46                    & 89.36          & 91.08                    & 91.03          & 76.56                    & 75.56                 & 76.66                    & 75.88                 & 77.66                    & 77.08                 \\
AMGCN \cite{wang2020gcn}        & 90.40                    & 90.43          & 90.76                    & 90.66          & 91.42                    & 91.36          & 81.89                    & 81.36                 & 84.94                    & 84.32                 & 87.30                    & 86.94                 \\ \hline
\textbf{SCRL} & \textbf{91.82}           & \textbf{91.79} & \textbf{92.06}           & \textbf{92.04} & \textbf{92.82}           & \textbf{92.80} & \textbf{90.22}           & \textbf{89.89}        & \textbf{90.26}           & \textbf{89.90}        & \textbf{91.58}           & \textbf{90.76}        \\ \hline
Dataset       & \multicolumn{6}{c|}{\textbf{Flickr}}                                                                                              & \multicolumn{6}{c|}{\textbf{UAI2010}}                                                                                                                      \\ \hline
L/C           & \multicolumn{2}{c|}{20}                   & \multicolumn{2}{c|}{40}                   & \multicolumn{2}{c|}{60}                   & \multicolumn{2}{c|}{20}                          & \multicolumn{2}{c|}{40}                          & \multicolumn{2}{c|}{60}                          \\ \hline
Label Rate           & \multicolumn{2}{c|}{2.38\%}                   & \multicolumn{2}{c|}{4.75\%}                   & \multicolumn{2}{c|}{7.13\%}                   & \multicolumn{2}{c|}{12.39\%}                          & \multicolumn{2}{c|}{24.78\%}                          & \multicolumn{2}{c|}{37.17\%}                          \\ \hline
Metrics       & \multicolumn{1}{c|}{ACC} & F1             & \multicolumn{1}{c|}{ACC} & F1             & \multicolumn{1}{c|}{ACC} & F1             & \multicolumn{1}{c|}{ACC} & F1                    & \multicolumn{1}{c|}{ACC} & F1                    & \multicolumn{1}{c|}{ACC} & F1                    \\ \hline
DeepWalk \cite{perozzi2014deepwalk}     & 24.33                    & 21.33          & 28.79                    & 26.90          & 30.10                    & 27.28          & 42.02                    & 32.92                 & 51.26                    & 46.01                 & 54.37                    & 44.43                 \\
LINE \cite{tang2015line}         & 33.25                    & 31.19          & 37.67                    & 37.12          & 38.54                    & 37.77          & 43.47                    & 37.01                 & 45.37                    & 39.62                 & 51.05                    & 43.76                 \\
ChebNet \cite{defferrard2016convolutional}     & 23.26                    & 21.27          & 35.10                    & 33.53          & 41.70                    & 40.17          & 50.02                    & 33.65                 & 58.18                    & 38.80                 & 59.82                    & 40.60                 \\
GCN \cite{kipf2016semi}          & 41.42                    & 39.95          & 45.48                    & 43.27          & 47.96                    & 46.58          & 49.88                    & 32.86                 & 51.80                    & 33.80                 & 54.40                    & 32.14                 \\
kNN-GCN \cite{wang2020gcn}       & 69.28                    & 70.33          & 75.08                    & 75.40          & 77.94                    & 77.97          & 66.06                    & 52.43                 & 68.74                    & 54.45                 & 71.64                    & 54.78                 \\
GAT \cite{velivckovic2017graph}          & 38.52                    & 37.00          & 38.44                    & 36.94          & 38.96                    & 37.35          & 56.92                    & 39.61                 & 63.74                    & 45.08                 & 68.44                    & 48.97                 \\
Demo-Net \cite{wu2019net}     & 34.89                    & 33.53          & 46.57                    & 45.23          & 57.30                    & 56.49          & 23.45                    & 16.82                 & 30.29                    & 26.36                 & 34.11                    & 29.03                 \\
MixHop \cite{abu2019mixhop}       & 39.56                    & 40.13          & 55.19                    & 56.25          & 64.96                    & 65.73          & 61.56                    & 49.19                 & 65.05                    & 53.86                 & 67.66                    & 56.31                 \\
GRACE \cite{Zhu2020vf}         & 49.42                    & 48.18          & 53.64                    & 52.61          & 55.67                    & 54.61          & 65.54                    & 48.38                 & 66.67                    & 49.50                 & 68.68                    & 51.51                 \\
AMGCN \cite{wang2020gcn}         & 75.26                    & 74.63          & 80.06                    & 79.36          & 82.10                    & 81.81          & 70.10                    & 55.61                 & 73.14                    & 64.88                 & 74.40                    & 65.99                 \\ \hline
\textbf{SCRL} & \textbf{79.52}           & \textbf{78.89} & \textbf{84.23}           & \textbf{84.03} & \textbf{84.54}           & \textbf{84.51} & \textbf{72.90}           & \textbf{57.80}        & \textbf{74.58}           & \textbf{67.40}        & \textbf{74.90}           & \textbf{67.54}        \\ \hline
Dataset       & \multicolumn{6}{c|}{\textbf{Citeseer}}                                                                                            & \multicolumn{6}{c|}{\textbf{PubMed}}                                                                                                                   \\ \hline
L/C           & \multicolumn{2}{c|}{20}                   & \multicolumn{2}{c|}{40}                   & \multicolumn{2}{c|}{60}                   & \multicolumn{2}{c|}{20}                          & \multicolumn{2}{c|}{40}                          & \multicolumn{2}{c|}{60}                          \\ \hline
Label Rate           & \multicolumn{2}{c|}{3.61\%}                   & \multicolumn{2}{c|}{7.21\%}                   & \multicolumn{2}{c|}{10.82\%}                   & \multicolumn{2}{c|}{0.30\%}                          & \multicolumn{2}{c|}{0.61\%}                          & \multicolumn{2}{c|}{0.91\%}                          \\ \hline
Metrics       & \multicolumn{1}{c|}{ACC} & F1             & \multicolumn{1}{c|}{ACC} & F1             & \multicolumn{1}{c|}{ACC} & F1             & \multicolumn{1}{c|}{ACC} & F1                    & \multicolumn{1}{c|}{ACC} & F1                    & \multicolumn{1}{c|}{ACC} & F1                    \\ \hline
DeepWalk \cite{perozzi2014deepwalk}      & 43.47                    & 38.09          & 45.15                    & 43.18          & 48.86                    & 48.01          & -                        & -                     & -                        & -                     & -                        & -                     \\
LINE \cite{tang2015line}        & 32.71                    & 31.75          & 33.32                    & 32.42          & 35.39                    & 34.37          & -                        & -                     & -                        & -                     & -                        & -                     \\
ChebNet \cite{defferrard2016convolutional}      & 69.80                    & 65.92          & 71.64                    & 68.31          & 73.26                    & 70.31          & 74.20                        & 73.51                     & 76.00                        & 74.92                     & 76.51 &75.83                     \\
GCN \cite{kipf2016semi}         & 70.30                    & 67.50          & 73.10                    & 69.70          & 74.48                    & 71.24          & 79.00                    & 78.45                 & 79.98                    & 79.17                 & 80.06                    & 79.65                 \\
kNN-GCN \cite{wang2020gcn}      & 61.35                    & 58.86          & 61.54                    & 59.33          & 62.38                    & 60.07          &  71.62                  & 71.92                 & 74.02                    & 74.09                 & 74.66                    &  75.18                \\
GAT \cite{velivckovic2017graph}          & 72.50                    & 68.14          & 73.04                    & 69.58          & 74.76                    & 71.60          & -                        & -                     & -                        & -                     & -                        & -                     \\
Demo-Net \cite{wu2019net}     & 69.50                    & 67.84          & 70.44                    & 66.97          & 71.86                    & 68.22          & -                        & -                     & -                        & -                     & -                        & -                     \\
MixHop \cite{abu2019mixhop}       & 71.40                    & 66.96          & 71.48                    & 67.40          & 72.16                    & 69.31          & -                        & -                     & -                        & -                     & -                        & -                     \\
GRACE \cite{Zhu2020vf}        & 71.70                    & 68.14          & 72.38                    & 68.74          & 74.20                    & 70.73          & 79.50     & \textbf{79.33}   & 80.32     & 79.64 &80.24      & 80.33 \\
AMGCN \cite{wang2020gcn}       & 73.10                    & 68.42          & 74.70                    & 69.81          & 75.56                    & 70.92          & 76.18                    & 76.86                 & 77.14                    & 77.04                & 77.74                     & 77.09                  \\ \hline
\textbf{SCRL} & \textbf{73.62}           & \textbf{69.78} & \textbf{75.08}           & \textbf{70.68} & \textbf{75.96}           & \textbf{72.84} & \textbf{79.62}           & 78.88        & \textbf{80.74}           & \textbf{80.24}        & \textbf{81.03}           & \textbf{80.55}        \\ \hline
\end{tabular}
\end{table*}


\begin{figure*}[]
    \centering
    \begin{subfigure}[b]{0.16\textwidth}
         \centering
         \includegraphics[width=\textwidth]{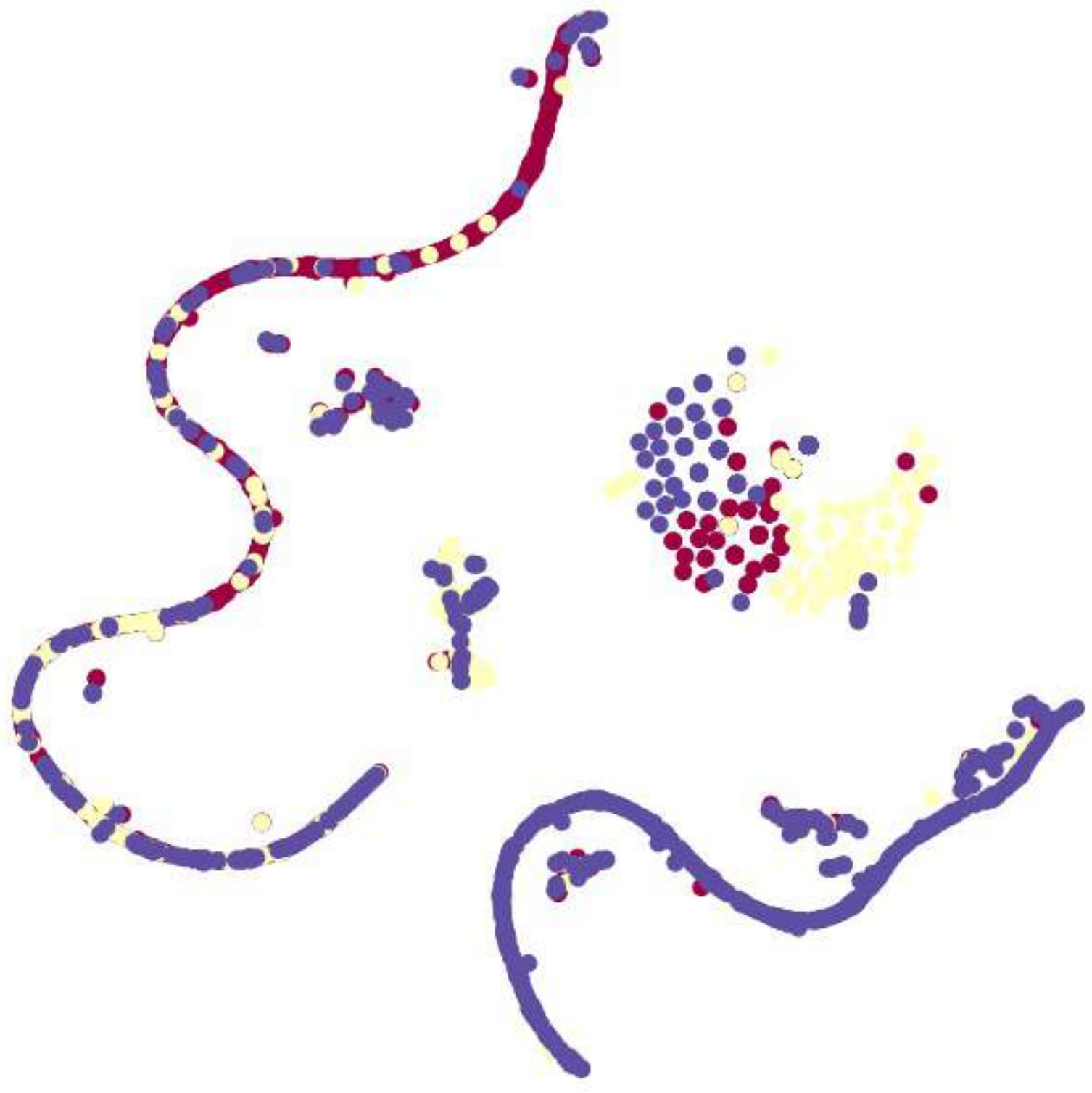}
         \caption{ACM Epoch 0}
         \label{fig: acmepoch0}
     \end{subfigure}
     \hfill
     \begin{subfigure}[b]{0.16\textwidth}
         \centering
         \includegraphics[width=\textwidth]{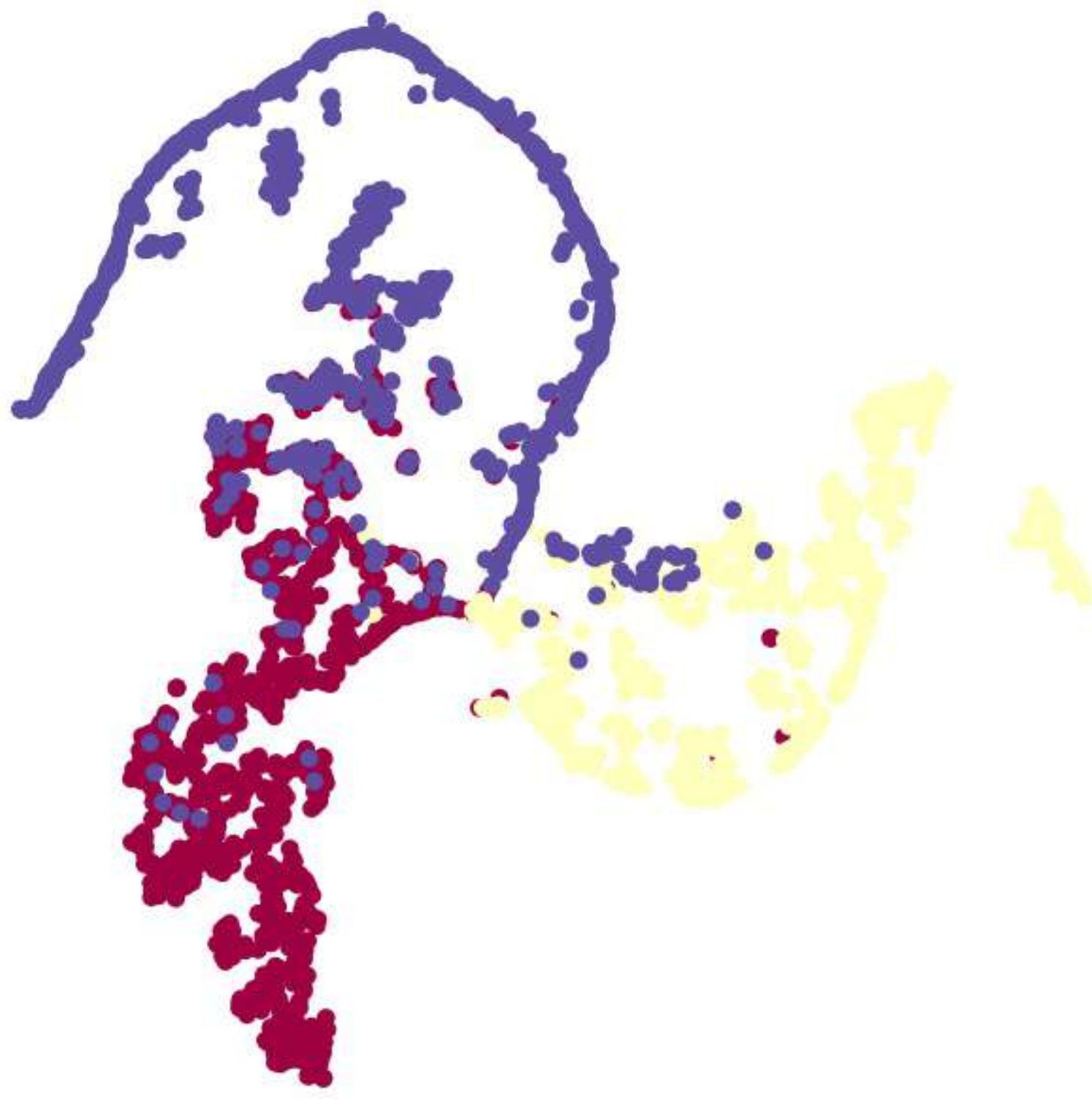}
         \caption{ACM Epoch 10}
         \label{gif: acmepoch10}
     \end{subfigure}
     \hfill
    \begin{subfigure}[b]{0.16\textwidth}
         \centering
         \includegraphics[width=\textwidth]{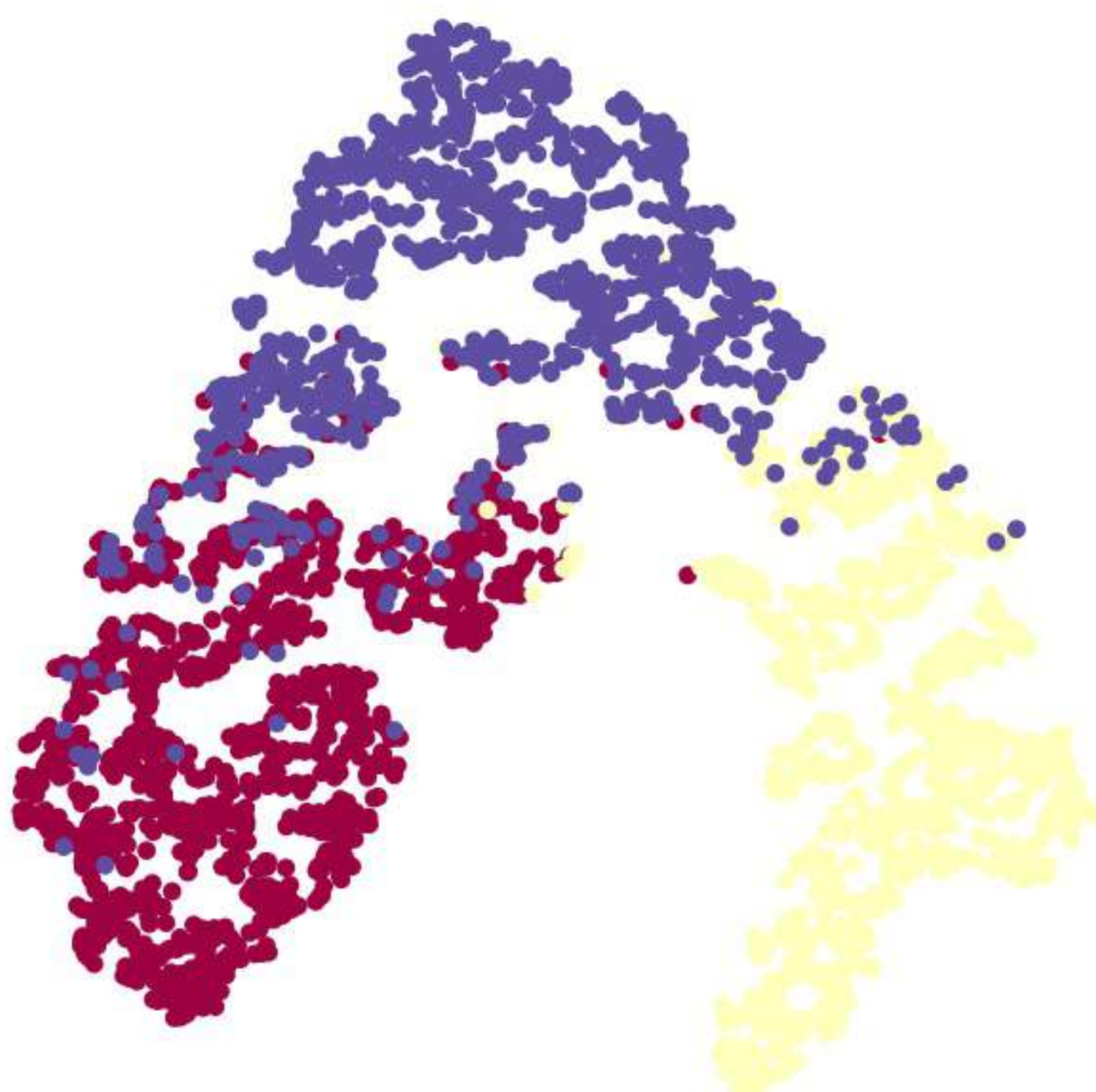}
         \caption{ACM Epoch 20}
         \label{fig: acmepoch20}
     \end{subfigure}
     \hfill
     \begin{subfigure}[b]{0.16\textwidth}
         \centering
         \includegraphics[width=\textwidth]{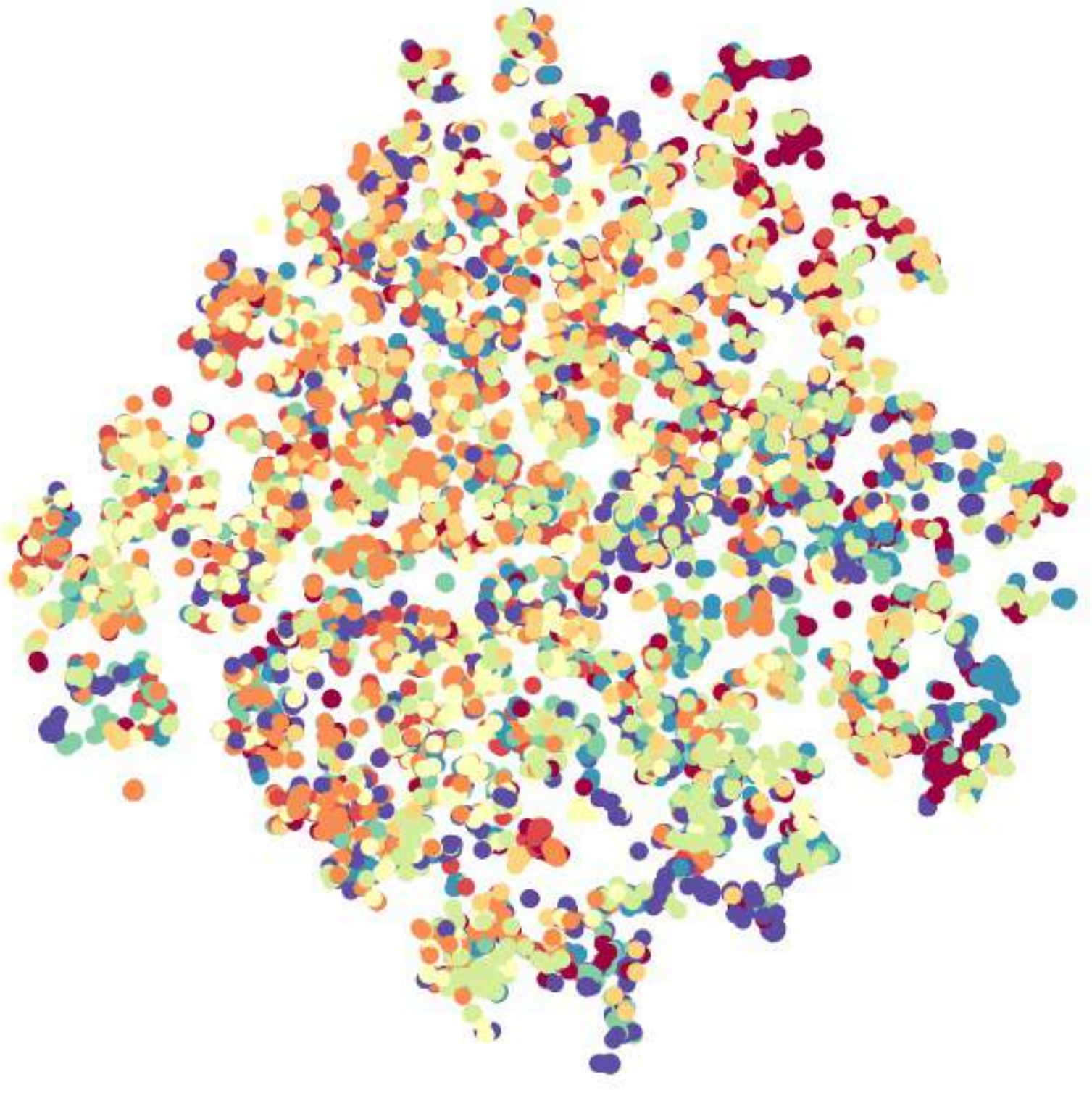}
         \caption{Flickr Epoch 0}
         \label{fig: flickrepoch0}
     \end{subfigure}
     \begin{subfigure}[b]{0.16\textwidth}
         \centering
         \includegraphics[width=\textwidth]{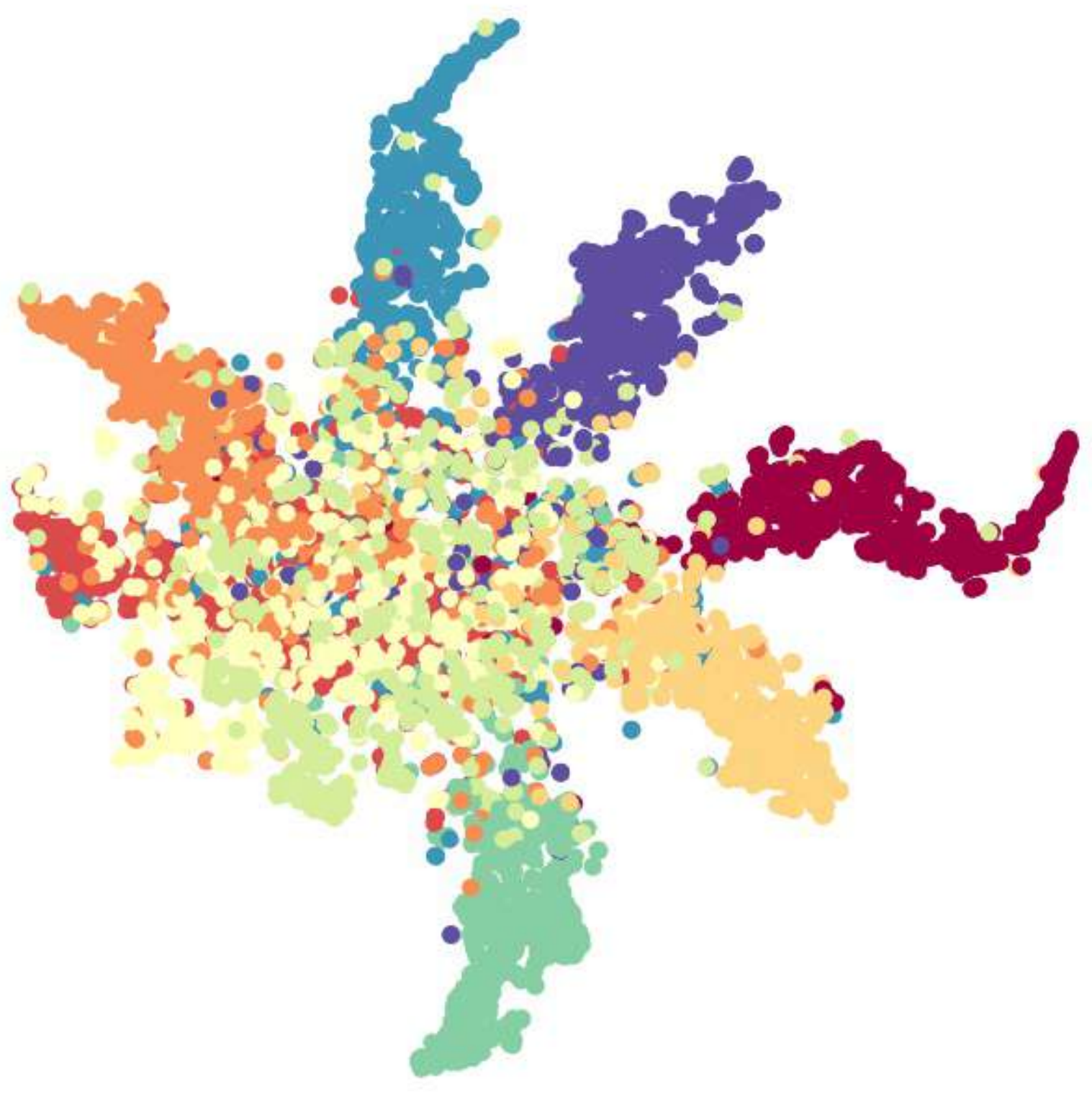}
         \caption{Flickr Epoch 30}
         \label{fig: flickrepoch30}
     \end{subfigure}
     \begin{subfigure}[b]{0.16\textwidth}
         \centering
         \includegraphics[width=\textwidth]{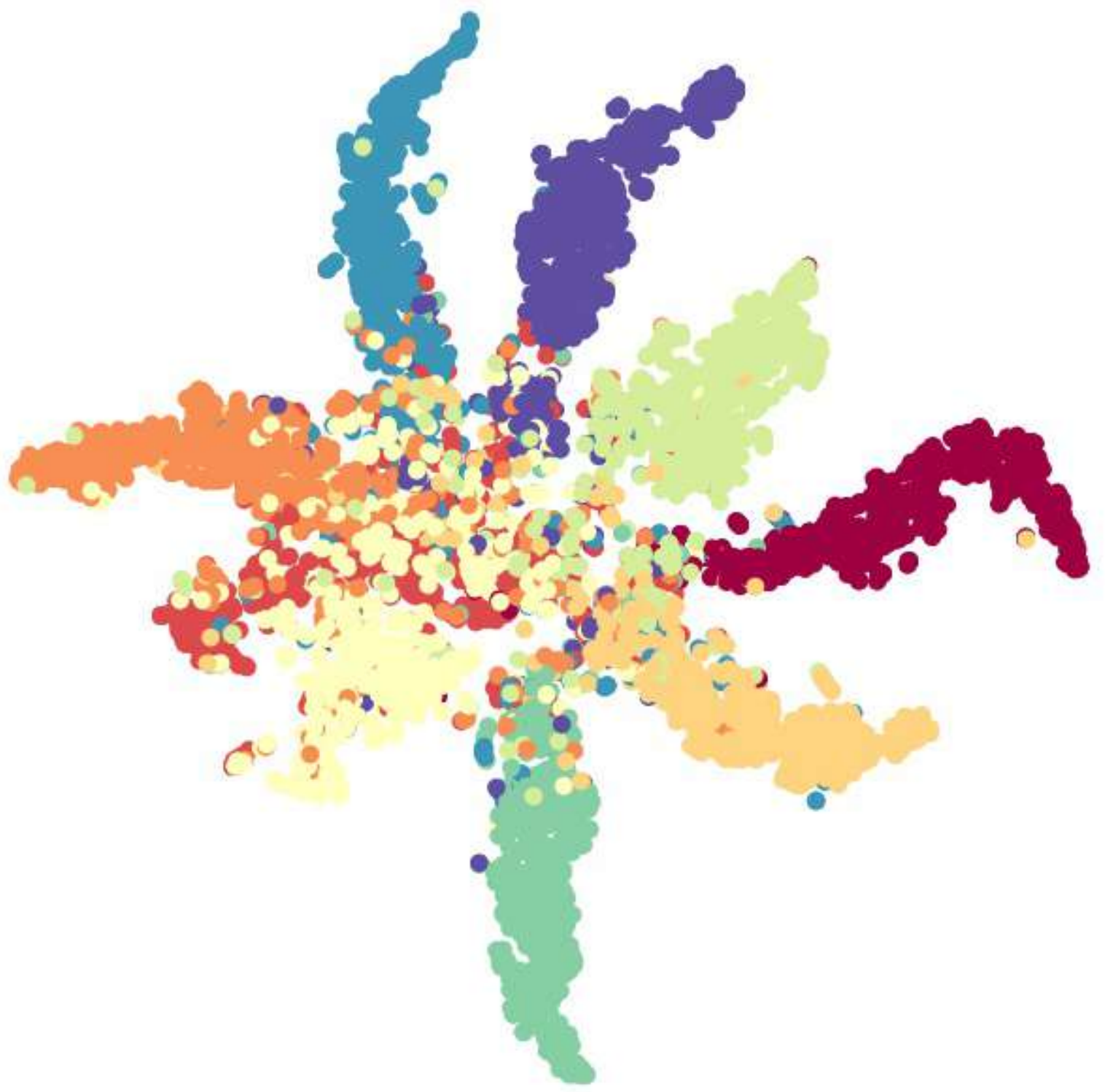}
         \caption{Flickr Epoch 60}
         \label{fig: flickrepoch60}
     \end{subfigure}
    \caption{The t-SNE demonstration of node representations of ACM and Flickr during training.}
    \label{fig:tsne graph during training}
\end{figure*}

\section{EXPERIMENT}
In this section, we conduct extensive experiments to evaluate the effectiveness of the 
self-supervised consensus representation learning framework for attributed graph.

\subsection{Datasets}
For the graph dataset, we select four commonly used citation networks (Citeseer \cite{kipf2016semi}, PubMed 
\cite{sen2008collective}, UAI2010 \cite{wang2018unified}, ACM \cite{wang2019heterogeneous})
 and two social networks (BlogCatalog \cite{meng2019co}, Flickr \cite{meng2019co}). 
 
 Specifically, Citeseer consists of 3327 scientific publications
 extracted from the CiteSeer digital library  classified into one of six classes. 
 PubMed consists of 19717 scientific publications from 
 PubMed database pertaining to diabetes classified into one of three classess.
 ACM network is
 extracted from ACM database where publications are represented by nodes and those
 with the same author are connected by edges. UAI2010 contains 
 3067 nodes in 19 classes and it has been tested in GCN
 for community detection. BlogCatalog
 is a social blog directory containing links between 5196 blogs. Flickr is 
 widely used by photo researchers and bloggers to host images that they embed 
 in blogs and social media. Flickr is composed of 7575 users and 
 they are classified into nine groups. The statistical information of datasets is summarized in Table \ref{table:dataset}.

\subsection{Baselines}
We thoroughly verify the performance of our proposed SCRL with representative baselines.

    DeepWalk \cite{perozzi2014deepwalk} is a graph embedding approach that 
    merely takes into account the structure of the graph.
    LINE \cite{tang2015line} is a graph embedding method for very large graph that utilizes both first-order and second-order proximity of the network.
    ChebNet \cite{defferrard2016convolutional} is a spectral-based GCN that
    uses Chebyshev polynomials to reduce computational complexity.
    GCN \cite{kipf2016semi} further solves the efficiency problem by 
    introducing first-order approximation of ChebNet. For comparison, we use the sparse k-nearest neighbor graph 
    calculated from feature matrix as the input graph of GCN and name it kNN-GCN.
    GAT \cite{velivckovic2017graph} adopts attention mechanism to learn the 
    relative weights between two connected nodes.
    Demo-Net \cite{wu2019net} is a degree-specific graph neural network
    for node classification.
    MixHop \cite{abu2019mixhop} is a GCN-based method that concatenates embeddings aggregated
    using the transition matrices of k-hop random walks before each layer.
    DGI \cite{velickovic2018deep} leverages local mutual information maximization across the
    graph’s patch representations.
    M3S \cite{sun2020multi} is a self-supervised framework that utilizes DeepCluster \cite{caron2018deep} to choose nodes with precise pseudo labels.
    GRACE \cite{Zhu2020vf} is a recently proposed graph contrastive 
    learning framework. It generates two graph views by corruption and learns node representations by 
    maximizing the agreement of node representations in these two views.
    AMGCN \cite{wang2020gcn} extracts embeddings from node features, topological structure, 
     and uses the attention mechanism to learn the adaptive importance weights of embeddings.

\subsection{Experimental Setup}
The experiments are run on the PyTorch platform using an Intel(R) Core(TM) i7-8700 CPU, 
64G RAM and GeForce GTX 1080 Ti 11G GPU.
Technically two layer GCN are built and we train
our model by utilizing the Adam \cite{kingma2014adam} optimizer with learning rate ranging
from 0.0001 to 0.0005. In order to prevent over-fitting, we set the dropout rate to 0.5.
In addition, we set weight decay $\in \left\{1e-4, \cdots, 5e-3 \right\}$ and k $\in \left\{2, \cdots, 9 \right\}$ for the kNN graphs. Two popular metrics are applied to quantitatively evaluate the semi-supervised node classification
performance: Accuracy (ACC) and F1-Score (F1). For fairness, we follow Wang \emph{et al.} \cite{wang2020gcn} and Yang \emph{et al.} \cite{yang2016revisiting} and 
select 20, 40, 60 nodes per class for training and 1000 nodes for testing.
For example, there are 19 types of nodes in UAI2010, therefore we train our model 
on training set with 380/760/1140 nodes, corresponding to label rate of 12.39\%, 24.78\%, 37.17\%, respectively. The selection of labeled nodes on each dataset is identical for all compared baselines.
We repeatedly train and test our model
for 5 times with the same partition of dataset and then report the average of ACC and F1.

\begin{figure}[t]
    \centering
    \begin{subfigure}[b]{0.11\textwidth}
         \centering
         \includegraphics[width=\textwidth]{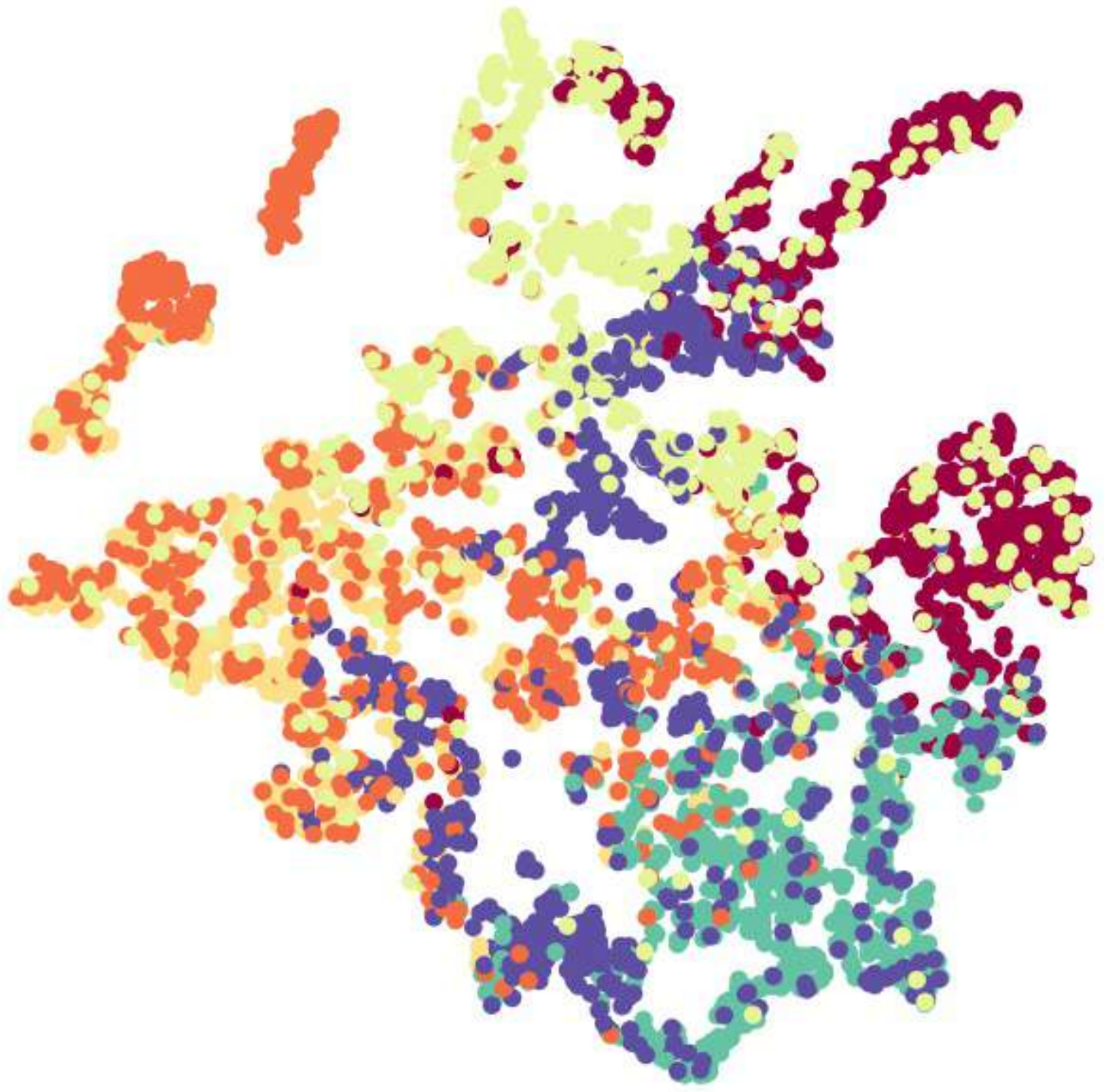}
         \caption{GCN}
         \label{fig: blog20gcn}
     \end{subfigure}
     \hfill
     \begin{subfigure}[b]{0.11\textwidth}
         \centering
         \includegraphics[width=\textwidth]{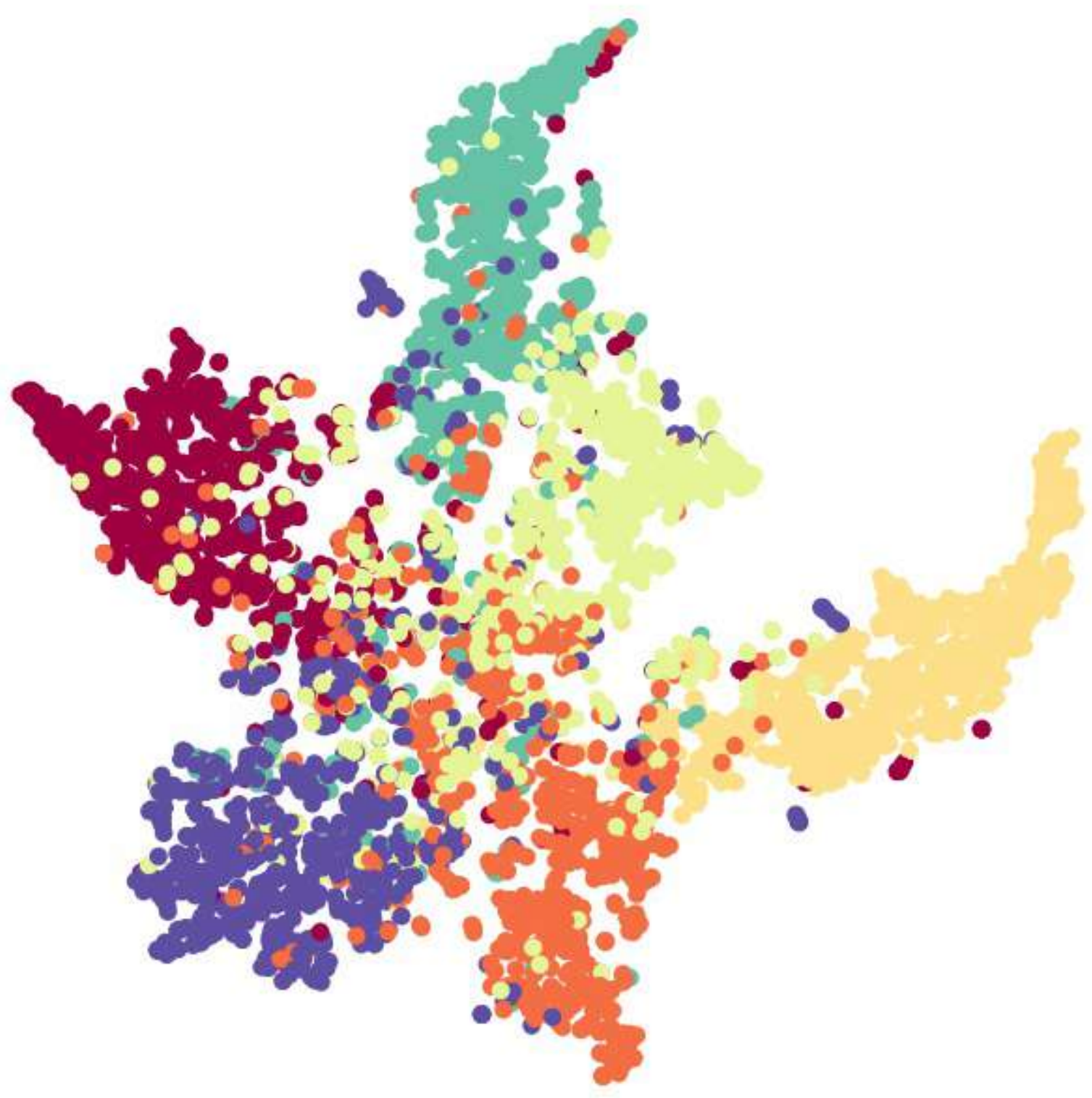}
         \caption{AMGCN}
         \label{gif: blog20amgcn}
     \end{subfigure}
     \hfill
    \begin{subfigure}[b]{0.11\textwidth}
         \centering
         \includegraphics[width=\textwidth]{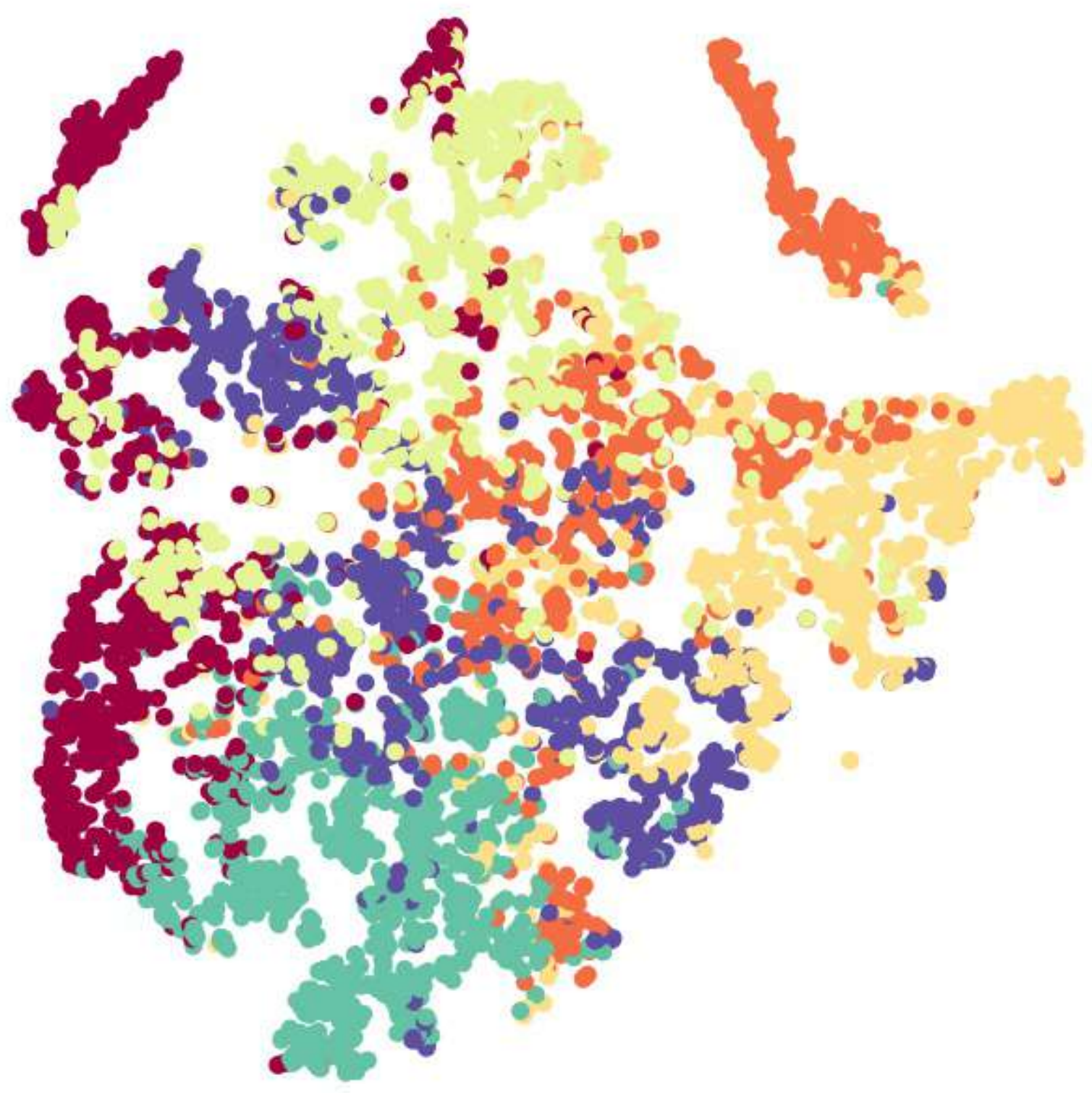}
         \caption{GRACE}
         \label{fig: blog20amgcn}
     \end{subfigure}
     \hfill
     \begin{subfigure}[b]{0.11\textwidth}
         \centering
         \includegraphics[width=\textwidth]{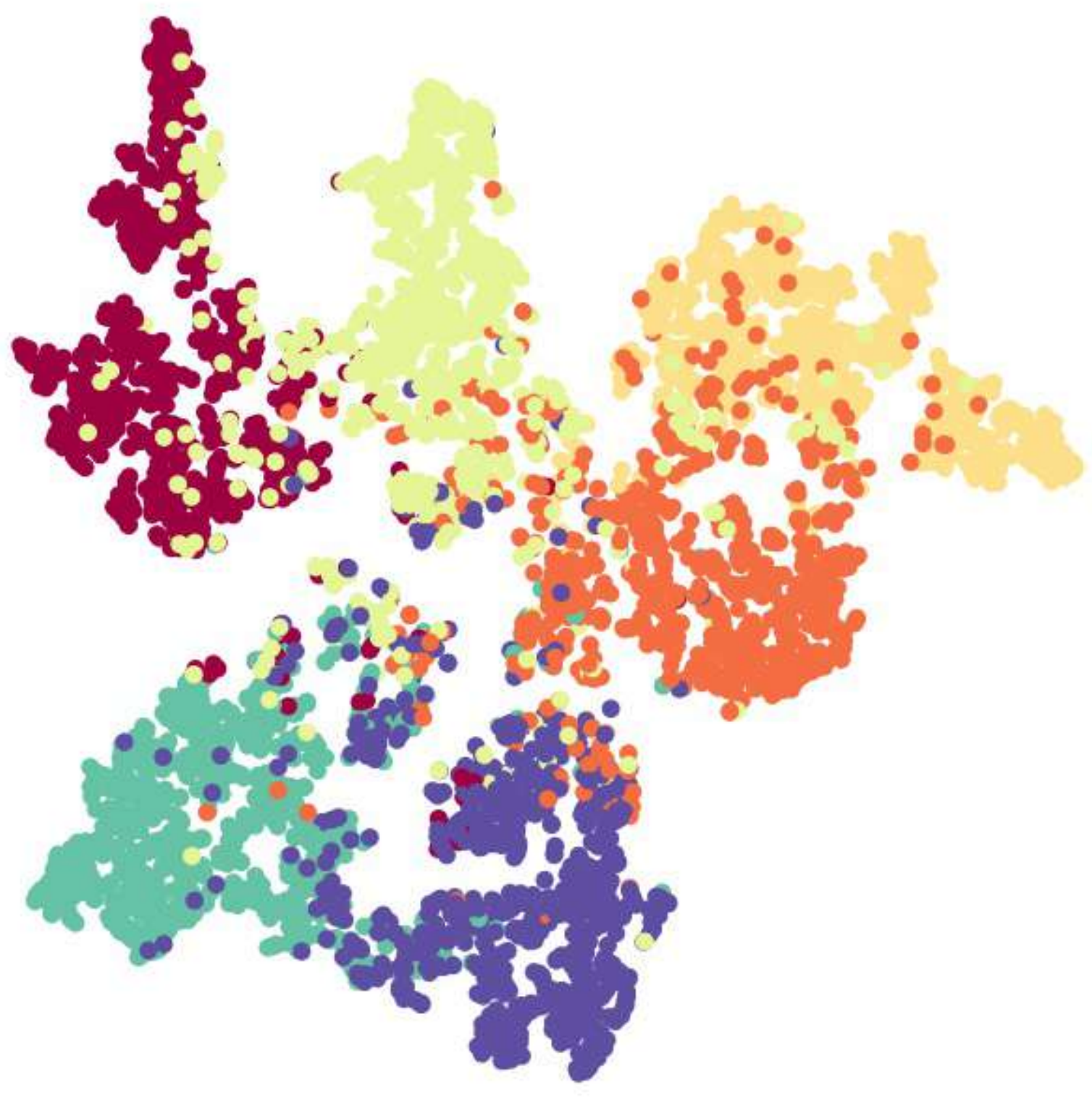}
         \caption{SCRL}
         \label{fig: blog20ls2gcn}
     \end{subfigure}
    \caption{Visualization of learnt representations of different methods on 
    BlogCatalog dataset.}
    \label{fig:tsne_blog}
\end{figure}

\begin{figure}[]
\centering
  \includegraphics[width=0.95\linewidth]{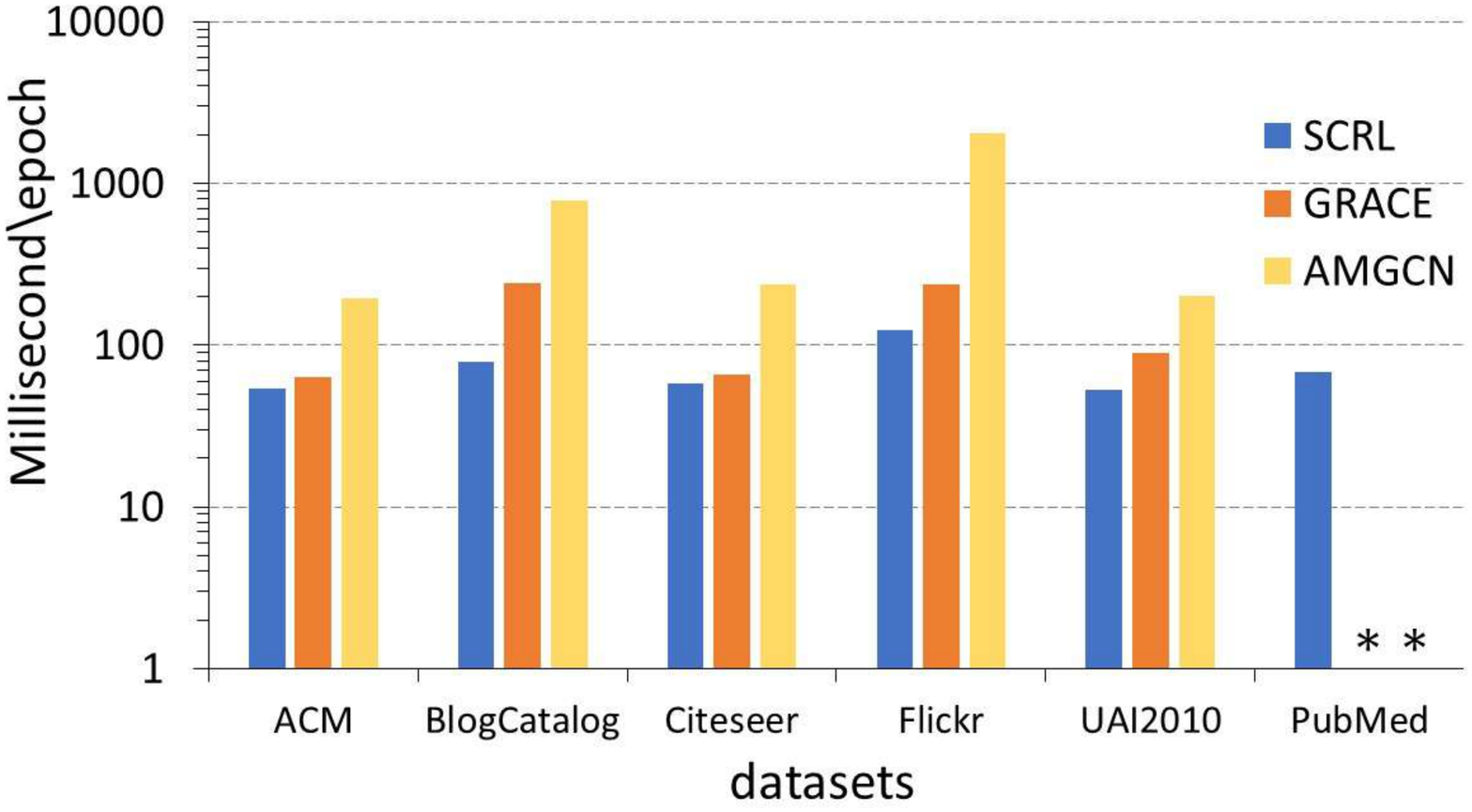}
  \caption{Averaged time cost per epoch of AMGCN, GRACE and SCRL for six datasets. 
  (*) indicates out-of-memory error and vertical axis is in log-scale.}
  \Description{Time per epoch}
  \label{fig: time}
\end{figure}

\subsection{Node Classification Results}


The results of experiments are summarized in Table \ref{table:result}, where the best performance is highlighted in bold.
Some results are directly taken from \cite{wang2020gcn}. We have following findings:
\begin{itemize}
    \item It can be seen that our proposed method boosts the performance
    of the listed baselines across most evaluation metrics on six datasets, which proves 
    its effectiveness. Particularly, compared with AMGCN, SCRL achieves a maximum improvement 
    of $8.33\%$ for ACC and $8.53\%$ for F1 on BlogCatalog.
    Additionally, on Flickr our method can exceed AMGCN by $4.26\%$ and $4.67\%$  
    for ACC and F1, respectively. This is mainly attributed to the self-supervision component.
    \item Our SCRL achieves better performances than GRACE on most of the metrics, 
    especially when the label 
    rate is relatively low. This could be explained by the fact that GRACE performs
    corruption by randomly adding/deleting edges to generate 
    views for graphs that may
    damage the original graph topology, hence degrading performance of classification.
     \item On some occasions, feature graph produces better result than topology graph.
    For example, on BlogCatalog, 
    Flickr, and UAI2010, kNN-CGN easily outperforms GCN.
    This confirms the necessity of incorporating the feature graph into our framework.
    On all datasets, SCRL achieves impressive 
    improvement compared with both GCN and kNN-GCN, which indicates that the 
    consensus representation of two graphs
    provides more holistic information than a single graph.
    \end{itemize}
    
    For a more intuitive understanding, 
    we use t-SNE to visualize the evolution of the representation learnt by our SCRL in the training
    process. As shown in Fig.\ref{fig:tsne graph during training}, at the beginning the 
    representation of ACM is
    chaotic and scattered. At apoch 10, compact clusters begin to form. By 
    epoch 20, a well learnt representation is established that makes it easier to separate
    data points into different groups. The representation of Flickr has a similar evolution process in the training.
    To further show the advantage of our proposed method, 
    we also visualize the embedding results of BlogCatalog generated by GCN, GRACE, AMGCN and 
    SCRL, which are shown in Fig.\ref{fig:tsne_blog}. It can be observed that
    the embedding generated by our proposed SCRL exhibits clearer
    cluster structure compared to other three methods.
   
    To verify the efficiency of SCRL, 
   we report the averaged training time per epoch when training SCRL, GRACE and AMGCN in Fig.\ref{fig: time}.
   Experiments are conducted with a GeForce GTX 1080 Ti 11G GPU.
    It can be seen that SCRL always costs much less time to train than others. For instance, for Flickr
    our proposed SCRL costs 65.9ms per epoch but AMGCN and GRACE need 237.7ms and 122.8ms, respectively.
    For BlogCatalog, SCRL needs 78.2ms per epoch while AMGCN needs 784.5ms and SCRL needs 243.8ms. 
    What's more, for larger datasets like PubMed, AMGCN and GRACE are subject to out-of-memory error.
    As mentioned above, AMGCN introduces attention mechanism and computes attention 
    weights for every node. GRACE divides node in two views into positive pairs,
    inter-view negative pairs and intra-view negative pairs, and then makes pair-wise comparisons in the 
    contrastive loss. Therefore, both AMGCN and GRACE become computationally inefficient and resource-consuming during training.

\begin{table}[]
\caption{Classification accuracy on Citeseer and PubMed with low label rates.}
\label{table: few_label}
\centering
\renewcommand{\arraystretch}{.9}
\begin{tabular}{c|cccc|ccc}
\hline
Datasets   & \multicolumn{4}{c|}{Citeseer}                                 & \multicolumn{3}{c}{PubMed}                    \\ \hline
L/C & 3 & 6 & 12 & 18 & 2 & 3 & 7 \\ \hline
Label Rate & 0.5\%         & 1\%           & 2\%           & 3\%           & 0.03\%        & 0.05\%        & 0.10\%        \\ \hline
ChebNet \cite{defferrard2016convolutional}    & 19.7          & 59.3          & 62.1          & 66.8          & 55.9          & 62.5          & 69.5          \\
GCN \cite{kipf2016semi}       & 33.4          & 46.5          & 62.6          & 66.9          & 61.8          & 68.8          & 71.9          \\
GAT \cite{velivckovic2017graph}       & 45.7          & 64.7          & 69.0          & 69.3          & 65.7          & 69.9          & 72.4          \\
DGI \cite{velickovic2018deep}      & 60.7          & 66.9          & 68.1          & 69.8          & 60.2          & 68.4          & 70.7          \\
M3S \cite{sun2020multi}    & 56.1          & 62.1          & 66.4          & 70.3          & 59.2          & 64.4          & 70.5          \\
GRACE \cite{Zhu2020vf}      & 55.4          & 59.3          & 63.4          & 67.8          & 64.4          & 67.5          & 72.3          \\
AMGCN \cite{wang2020gcn}     & 60.2          & 65.7          & 68.5          & 70.2          & 60.5              &  62.4             & 70.8              \\ \hline
SCRL       & \textbf{62.4} & \textbf{67.3} & \textbf{69.8} & \textbf{73.3} & \textbf{67.9} & \textbf{71.9} & \textbf{73.4} \\ \hline
\end{tabular}
\end{table}

\subsection{Few Labeled Classification}
To further investigate the capability of our proposed SCRL in dealing with scarce supervision data,
we conduct experiments when the number of labeled examples is extremely small.
Taking Citeseer and PubMed for example, we strictly follow Li \emph{et al.} \cite{li2018deeper} and 
select a small set of labeled examples for model training. 
Specifically, for Citeseer, we select 3, 6, 12, 18 nodes per class, corresponding to 
four label rates: $0.5\%$, $1\%$, $2\%$ and $3\%$.
For PubMed, we select 2, 3, 7 nodes per class, corresponding to three label rates:
$0.03\%$, $0.05\%$ and $0.10\%$. To
make a fair comparison, we report mean 
classification accuracy of 10 runs.

We report the result in Table \ref{table: few_label}.
We can observe that SCRL outperforms all state-of-the-art approaches.
It can be seen that the accuracy of GCN, ChebNet, and GAT decline severely when the label rate is very low, 
especially on $0.5\%$ Citeseer, due to insufficient propagation of label information.
By contrast, self-supervised/contrastive approaches, i.e., DGI, M3S, GRACE, 
are obviously much better because they additionally exploit supervisory 
signals with data themselves. Different from them, SCRL explores supervisory 
signals in feature graph, which is ignored by other self-supervised methods.
Specifically, SCRL improves DGI, M3S, GRACE by $3.03\%$, $5.29\%$, and $5.12\%$ on average, respectively.

\begin{figure}
    \centering
    \begin{subfigure}[b]{0.45\linewidth}
         \centering
         \includegraphics[width=\textwidth]{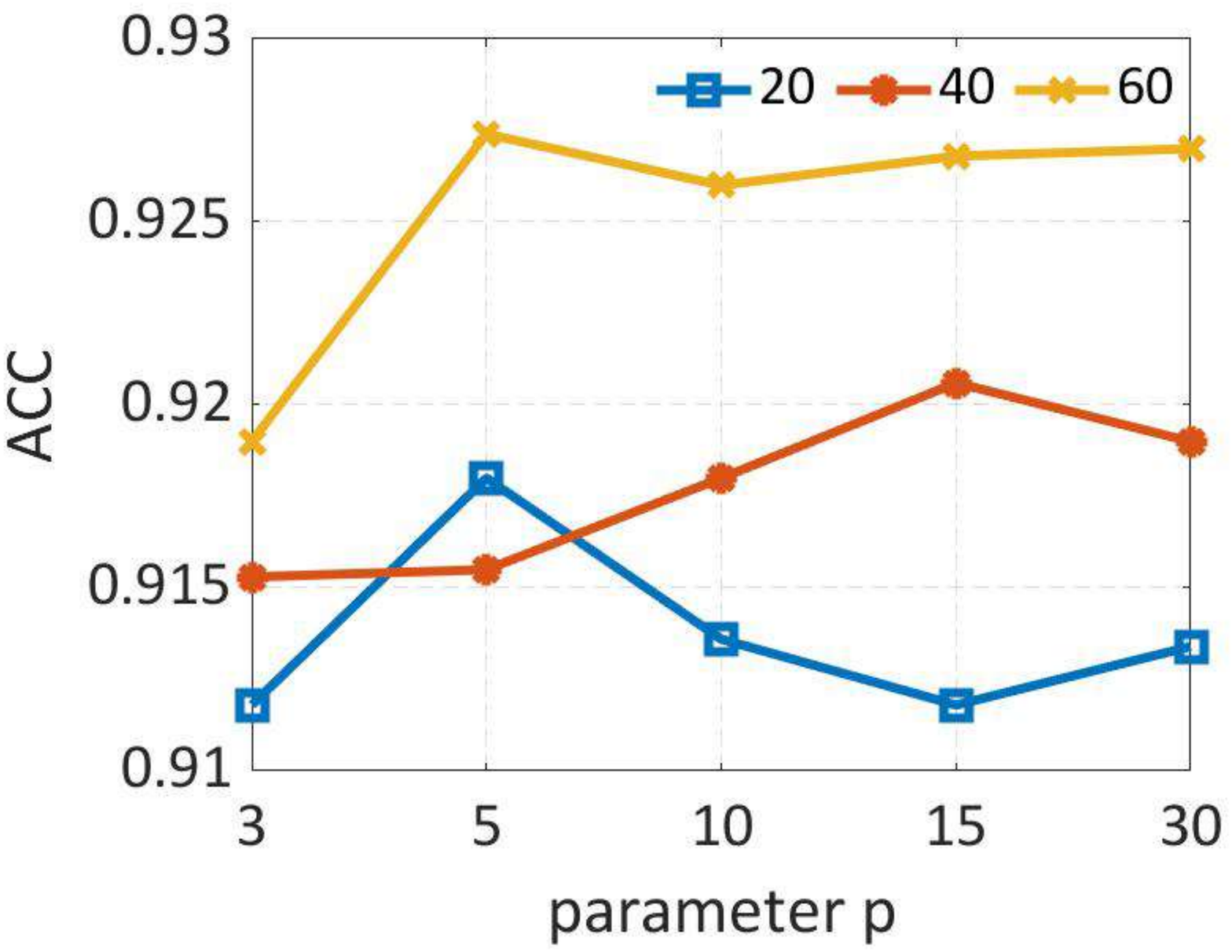}
         \caption{ACM}
         \label{ACM}
     \end{subfigure}
     \hfill
     \begin{subfigure}[b]{0.45\linewidth}
         \centering
         \includegraphics[width=\textwidth]{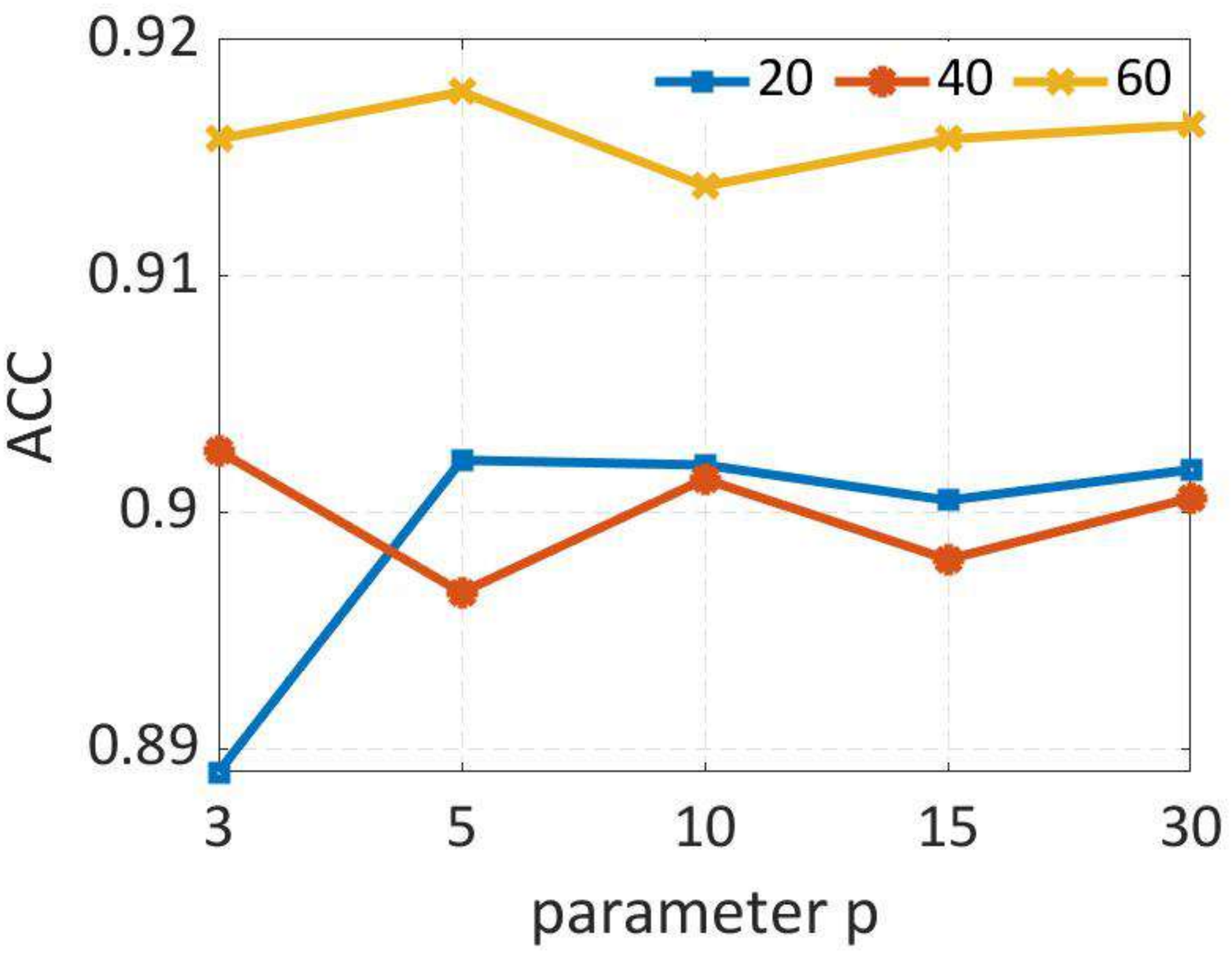}
         \caption{BlogCatalog}
         \label{BlogCatalog}
     \end{subfigure}
     \hfill
     \begin{subfigure}[b]{0.45\linewidth}
         \centering
         \includegraphics[width=\textwidth]{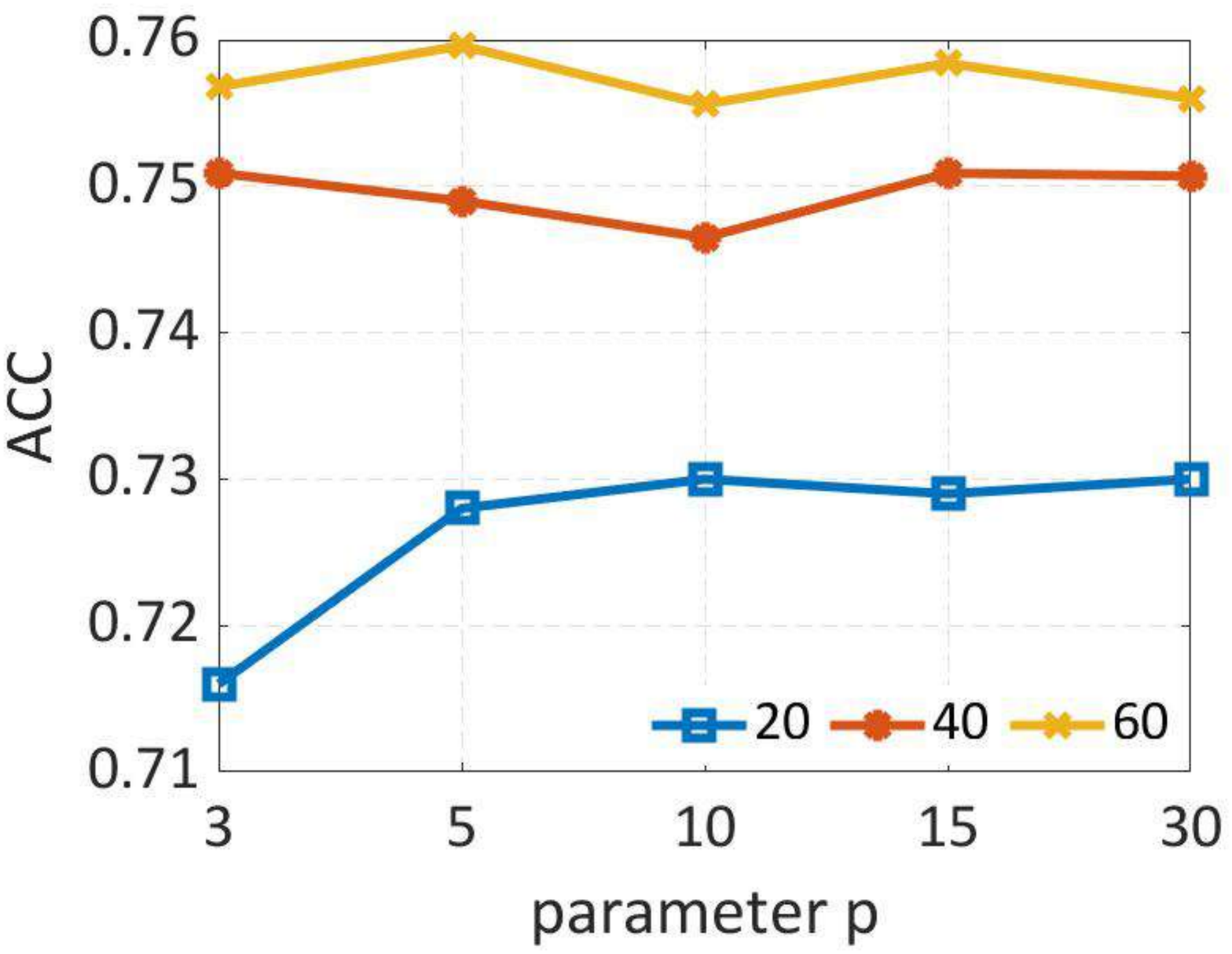}
         \caption{Citeseer}
         \label{Citeseer}
     \end{subfigure}
     \hfill
     \begin{subfigure}[b]{0.45\linewidth}
         \centering
         \includegraphics[width=\textwidth]{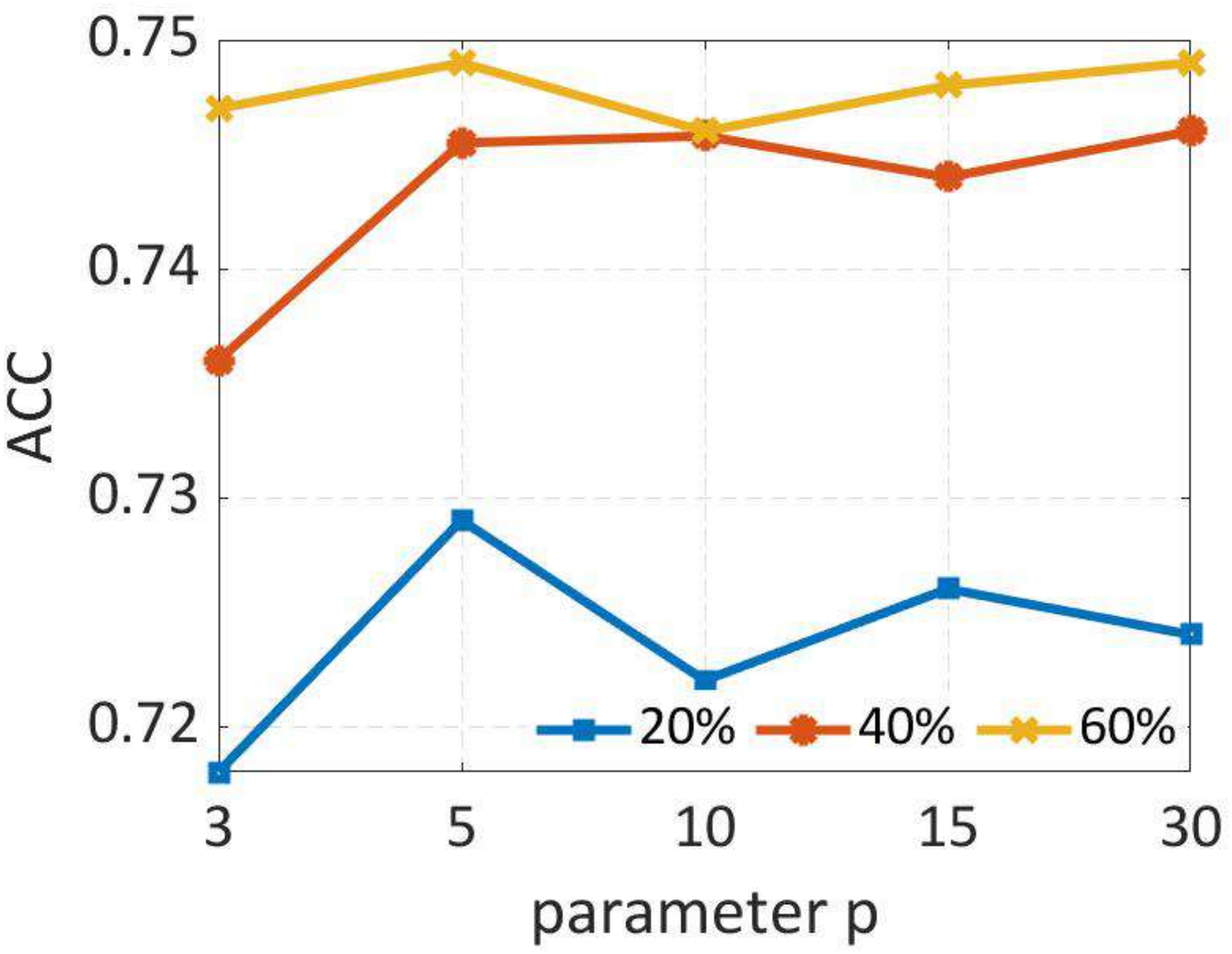}
         \caption{UAI2010}
         \label{UAI}
     \end{subfigure}
    \caption{The influence of iteration number in Sinkhorn algorithm.}
    \label{fig: iterate}
\end{figure}

\begin{figure}[h]
   \centering
    \begin{subfigure}[b]{0.48\linewidth}
         \centering
         \includegraphics[width=\textwidth]{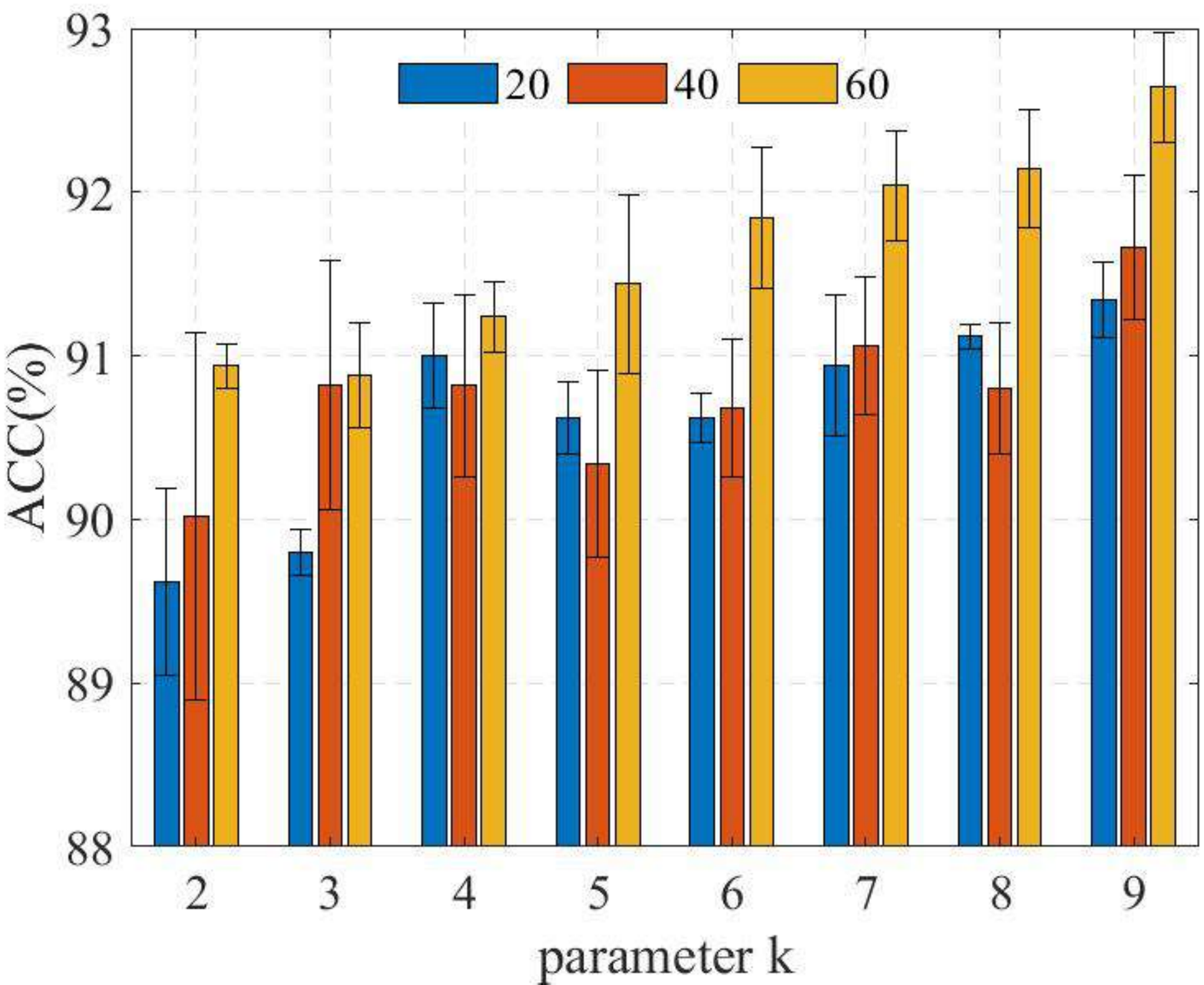}
         \caption{ACM}
         \label{ACM}
     \end{subfigure}
     \hfill
    \centering
    \begin{subfigure}[b]{0.48\linewidth}
         \centering
         \includegraphics[width=\textwidth]{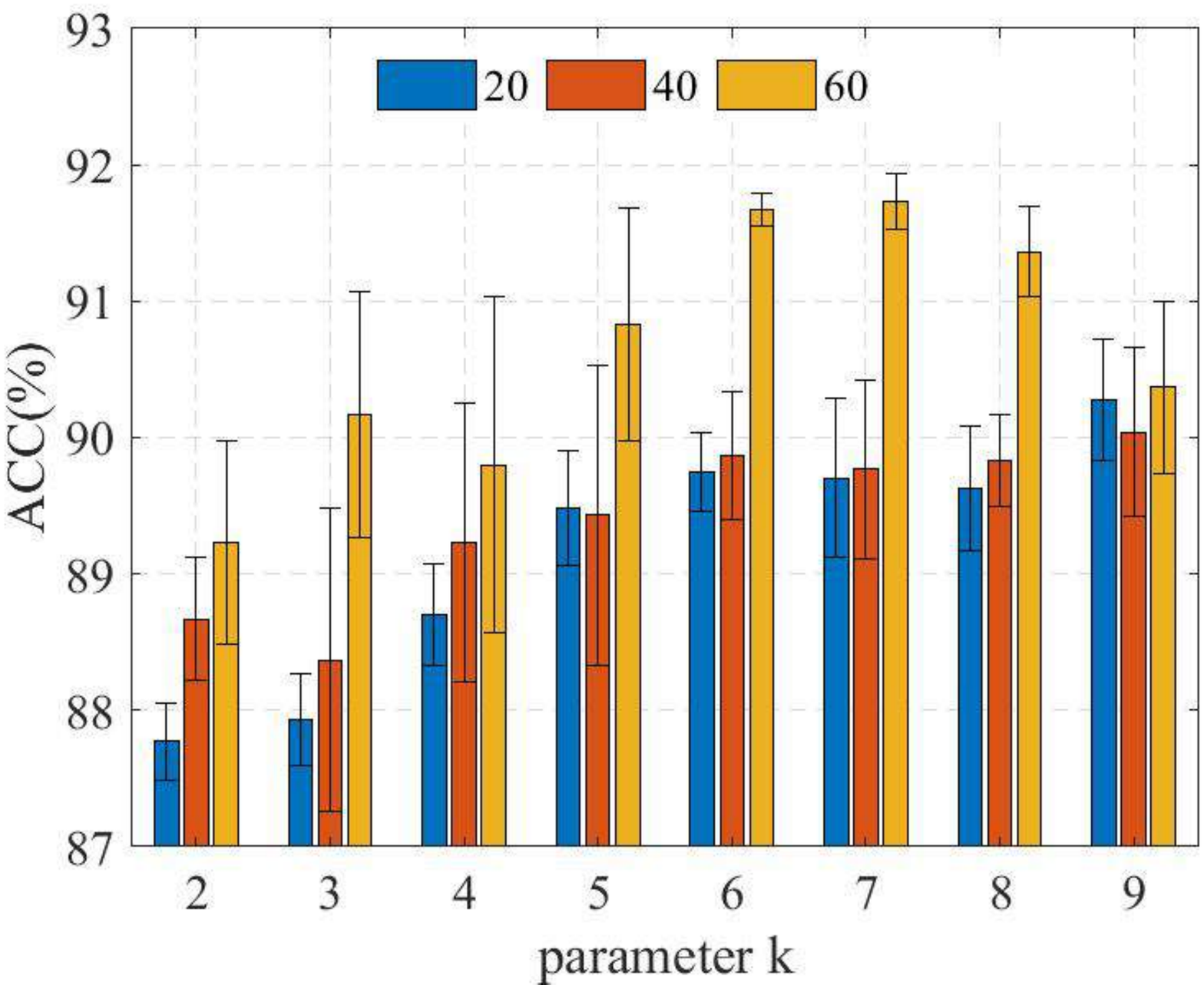}
         \caption{BlogCatalog}
         \label{ACM}
     \end{subfigure}
  \caption{The influence of parameter k in feature graph.}
  \label{fig:para k}
  \Description{analysis of parameter k}
\end{figure}

\subsection{Parameter Analysis}
In this section, we analyze the sensitivity of parameters of our method on ACM, BlogCatalog, Citeseer, and UAI2020.
\begin{table}[]
\caption{The influence of prototype number on Flickr.}
\label{table: num_pro}
\begin{tabular}{c|cccc}
\hline
Number of prototypes & 9(M)     & 27(3M)    & 45(5M)    & 90(10M)    \\ \hline
20 L/C               & 78.54 & \textbf{79.52} & 78.82 & 78.75 \\
40 L/C               & 83.82 & 84.15 & \textbf{84.23} & 83.90 \\
60 L/C               & 84.03 & \textbf{84.54} & 84.48 & 84.42 \\ \hline
\end{tabular}
\end{table}

\subsubsection*{\textbf{Number of prototypes $B$}}
In Table \ref{table: num_pro}, we evaluate the influence of prototype number on Flickr 
via varying the value of $B$.
We can observe that when $B$ is bigger than 27, 
increasing prototype number may not improve the performance substantially.
This suggests that $B$ has little influence as long as there are "enough" prototypes.
In fact, using too many prototypes increases computation time. Throughout this paper, we set $B$ in the range of $M$ to $3M$ when we train SCRL. 

\subsubsection*{\textbf{Number of iterations p}}
We investigate the impact of normalization steps performed during
Sinkhorn algorithm \cite{cuturi2013sinkhorn}. 
We test it by setting p to 5 values: 3, 5, 10, 15, 30.
The results are shown in Fig.\ref{fig: iterate}. As the number of iterations increases, the accuracy normally raises 
first and then holds steady. We observe that 5 iterations are usually good enough for the model
to reach an ideal accuracy.

\subsubsection*{\textbf{Parameter k}}
Finally, we study the impact of parameter k in the kNN feature graph
with various k ranging from 2 to 9. We conduct experiments on ACM and BlogCatalog and fix 
the iteration number p to 5.
For every k, we repeat experiments 10 times and record ACC to 
further calculate its average and standard deviation, which are displayed in Fig.\ref{fig:para k}.
For ACM, with the increasing of k, the performance usually becomes better as well. It is because a
larger k can provide more information about the relationship between nodes in feature space. 
Generally, BlogCatalog has a similar trend.
However, with 60 labeled nodes per class, SCRL achieves the best performance when k is 7. 
That is perhaps because too many neighbor nodes in feature graph may 
introduce some noisy edges.

\section{Ablation Study}
As mentioned above, our proposed SCRL model employs self-supervised loss to learn a consensus representation
from topological structure and node features. To elucidate the contribution of self-supervision module, we
report the classification results of SCRL when this component is removed on three datasets: 
Citeseer, ACM, and Flickr. For fairness, the split of data is identical with the experimental
setting in Section 4. We adopt "SCRL(w/o SSL)" to represent the simplified model when self-supervised loss (SSL)
is removed. The comparison is shown in Table \ref{table:ablation}.
Apparently, ACC and F1 decrease when the aforementioned component
is dropped from the framework. It reveals that our proposed self-supervision module is able to improve the performance
on semi-supervised learning task substantially. For example, the accuracy can be boosted by over $5\%$ in some cases
due to the introduction of self-supervised loss. In addition, the improvement of SCRL(sw/o SSL) over GCN verifies the 
importance of feature graph.


\begin{table}[]
\caption{The influence of self-supervised module.}
\label{table:ablation}
\renewcommand{\arraystretch}{.9}
\begin{tabular}{c|c|c|c|c}
\hline
Dataset & Metrics & L/C & SCRL  & \begin{tabular}[c]{@{}c@{}}SCRL\\ (w/o SSL)\end{tabular} \\ \hline
\multirow{6}{*}{Flickr}   & \multirow{3}{*}{ACC} & 20 & 79.52 & 76.28 \\ \cline{4-5} 
                          &                      & 40 & 84.23 & 79.10 \\ \cline{4-5} 
                          &                      & 60 & 84.54 & 83.58 \\ \cline{2-5} 
                          & \multirow{3}{*}{F1}  & 20 & 78.89 & 74.70 \\ \cline{4-5} 
                          &                      & 40 & 84.03 & 78.37 \\ \cline{4-5} 
                          &                      & 60 & 84.51 & 83.12 \\ \hline
\multirow{6}{*}{ACM}      & \multirow{3}{*}{ACC} & 20 & 91.82 & 90.24 \\ \cline{4-5} 
                          &                      & 40 & 92.06 & 90.32 \\ \cline{4-5} 
                          &                      & 60 & 92.82 & 91.50 \\ \cline{2-5} 
                          & \multirow{3}{*}{F1}  & 20 & 91.79 & 90.20 \\ \cline{4-5} 
                          &                      & 40 & 92.04 & 90.28 \\ \cline{4-5} 
                          &                      & 60 & 92.80 & 91.45 \\ \hline
\multirow{6}{*}{Citeseer} & \multirow{3}{*}{ACC} & 20 & 73.50  & 71.82 \\ \cline{4-5} 
                          &                      & 40 & 75.08 & 74.48 \\ \cline{4-5} 
                          &                      & 60 & 75.96 & 74.30 \\ \cline{2-5} 
                          & \multirow{3}{*}{F1}  & 20 & 69.91 & 68.26 \\ \cline{4-5} 
                          &                      & 40 & 70.41 & 69.35 \\ \cline{4-5} 
                          &                      & 60 & 72.81 & 70.74 \\ \hline
\end{tabular}
\end{table}

\section{Conclusion}
In this paper, we propose a self-supervised consensus representation learning framework  
for semi-supervised classification on attributed graph. We make the first attempt
to introduce the idea of self-supervised learning to integrate the correlated information 
from the topology structure and node features. Specifically, we require that the embeddings of feature graph and topology graph should be consistent and generate the same labels. It is realized by the "exchanged prediction" in the self-supervised module. Extensive experiments well demonstrate its superior performance over the state-of-the-art models on real world datasets.
For example, our SCRL improves AMGCN by $2.74\%$ and $2.94\%$ for ACC and F1 on average across all datasets; SCRL improves GRACE by $9.28\%$ and $10.58\%$ for ACC and F1.
\begin{acks}
This paper was in part supported by Grants from the Natural
Science Foundation of China (Nos. 61806045,U19A2059). \\

\end{acks}


\bibliographystyle{ACM-Reference-Format}
\bibliography{sample-base}


\end{document}